\documentclass[sigconf]{acmart}
\usepackage{enumitem}
\usepackage{amsthm}
\usepackage{amsmath}
\usepackage{graphicx}
\usepackage{multirow}
\usepackage{tabularx}
\usepackage{subfigure}
\usepackage{pifont}
\usepackage{color}

\AtBeginDocument{%
  }

\setcopyright{acmlicensed}
\copyrightyear{2026}
\acmYear{2026}
\acmDOI{XXXXXXX.XXXXXXX}
\acmConference[The ACM Web Conference '26]{The Web Conference 2026}{April 13--17,
  2026}{Dubai, United Arab Emirates}
\acmISBN{978-1-4503-XXXX-X/2018/06}

\begin{document}

\title{FairGE: Fairness-Aware Graph Encoding in Incomplete Social Networks}

\author{Renqiang Luo}
\affiliation{%
    \institution{Jilin University}
    \city{Changchun}
    \country{China}
}
\email{lrenqiang@outlook.com}

\author{Huafei Huang}
\affiliation{%
    \institution{Adelaide University}
    \city{Adelaide}
    \country{Australia}
    }
\email{hhuafei@outlook.com}

\author{Tao Tang}
\affiliation{%
    \institution{Zhejiang University of Technology}
    \city{Hangzhou}
    \country{China}
}
\email{tao.tang@ieee.org}

\author{Jing Ren}
\affiliation{%
    \institution{RMIT Univeristy}
    \city{Melbourne}
    \country{Australia}
}
\email{jing.ren@ieee.org}

\author{Ziqi Xu}
\authornote{Corresponding author.}
\affiliation{%
    \institution{RMIT University}
    \city{Melbourne}
    \country{Australia}
}
\email{ziqi.xu@rmit.edu.au}

\author{Mingliang Hou}
\affiliation{%
    \institution{Jinan University \& TAL Education Group}
    \city{Guangzhou}
    \country{China}
}
\email{teemohold@outlook.com}

\author{Enyan Dai}
\affiliation{%
    \institution{HKUST}
    \city{Guangzhou}
    \country{China}
}
\email{enyandai@hkust-gz.edu.cn}

\author{Feng Xia}
\affiliation{%
  \institution{RMIT University}
  \city{Melbourne}
  \country{Australia}
}
\email{f.xia@ieee.org}

\renewcommand{\shortauthors}{Luo et al.}



\begin{abstract}
Graph Transformers (GTs) are increasingly applied to social network analysis, yet their deployment is often constrained by fairness concerns.
This issue is particularly critical in incomplete social networks, where sensitive attributes are frequently missing due to privacy and ethical restrictions.
Existing solutions commonly generate these incomplete attributes, which may introduce additional biases and further compromise user privacy.
To address this challenge, FairGE (Fair Graph Encoding) is introduced as a fairness-aware framework for GTs in incomplete social networks.
Instead of generating sensitive attributes, FairGE encodes fairness directly through spectral graph theory. 
By leveraging the principal eigenvector to represent structural information and padding incomplete sensitive attributes with zeros to maintain independence, FairGE ensures fairness without data reconstruction. 
Theoretical analysis demonstrates that the method suppresses the influence of non-principal spectral components, thereby enhancing fairness. 
Extensive experiments on seven real-world social network datasets confirm that FairGE achieves at least a 16\% improvement in both statistical parity and equality of opportunity compared with state-of-the-art baselines. 
The source code is shown in~\url{https://github.com/LuoRenqiang/FairGE}.
\end{abstract}




\maketitle

\section{Introduction}

\begin{figure}[t]
    \centering
    \includegraphics[width=0.45\textwidth]{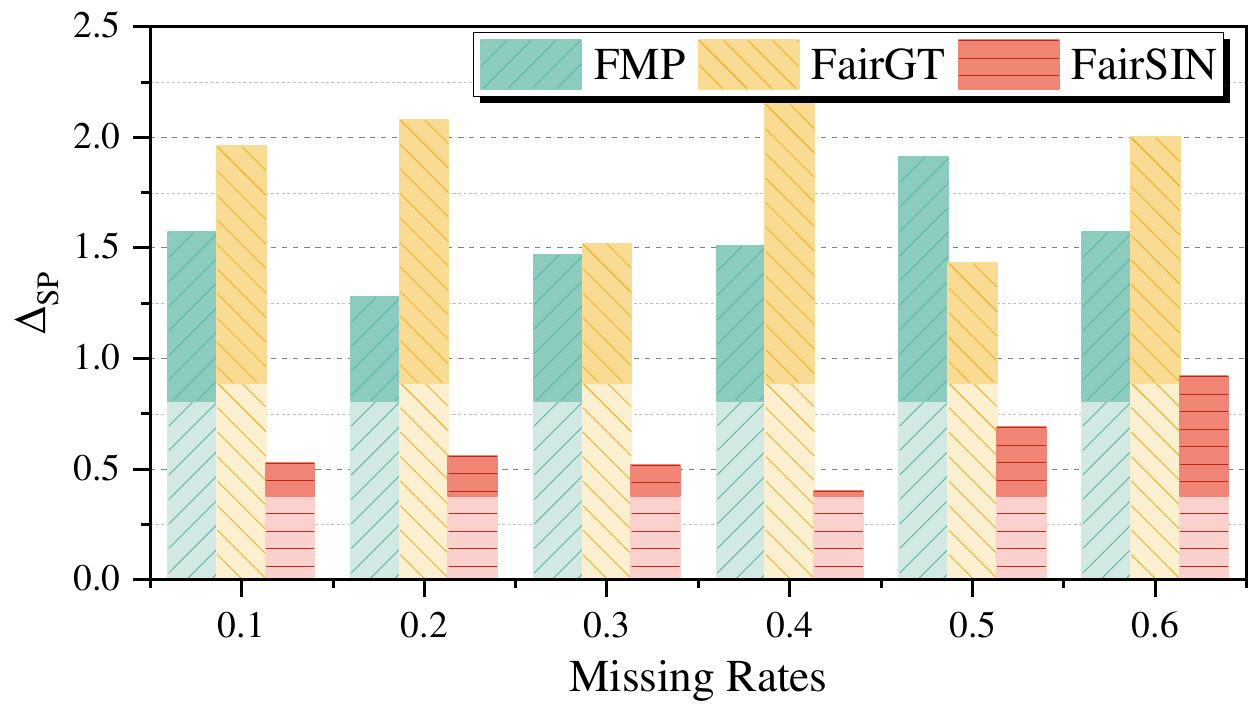}
    \caption{Impact of different missing rates of sensitive attributes on fairness-aware methods (FMP~\cite{jiang2024chasing}, FairGT~\cite{luo2024fairgt}, and FairSIN~\cite{yang2024fairsin}) on the Credit dataset~\cite{yeh2009the}.
    $\Delta_{\text{SP}}$ is reported as the fairness metric (see Section~\ref{Metrics}), where higher values indicate lower fairness. 
    Lighter bars correspond to results on complete data, while darker bars indicate the additional increase due to missing attributes. 
    For instance, in FMP, $\Delta_{\text{SP}}$ is about 0.78 with complete data but increases to around 1.6 at a missing rate of 0.1.}
    \label{fig:back}
\end{figure}

\par Graph Transformers (GTs) address common Graph Neural Networks (GNNs) challenges such as over-smoothing~\cite{zhou2025tokenphormer}, over-squash-ing~\cite{gravina2025on}, and long-range dependencies~\cite{chen2024masked} by employing global attention to capture long-range node interactions~\cite{wu2024on, zhang2025survey}. 
Despite these advantages, GTs often overlook inherent biases in graph data, leading to discriminatory predictions towards sensitive subgroups (e.g., gender and race)~\cite{luo2024fairgt}. 
This issue is particularly critical in social networks, where fairness directly affects user trust and system reliability~\cite{zhong2024adaptive, bernard2025a,abs-2504-21296}. 
Here, algorithmic fairness requires that model predictions remain unbiased with respect to sensitive attributes (e.g., gender, race, region, and other potential factors), ensuring equitable treatment of diverse subgroups~\cite{dong2023fairness,XuXLCLLW22,XuKON25,Oldfield0K25}. 
The challenge is further compounded by the frequent absence of sensitive attributes, as individuals are often unwilling or unable to disclose such information due to privacy and ethical concerns~\cite{dai2021say,XuCLLLW23}. 
Therefore, addressing fairness under incomplete sensitive attributes is essential for real-world social network applications, where reliable and equitable outcomes must be guaranteed~\cite{zhang2024heterogeneous}.

\par When under the incomplete social networks (i.e., sensitive attributes are partially missing), the fairness performance of existing fairness-aware methods often deteriorates. 
As shown in Figure~\ref{fig:back}, fairness degradation does not increase monotonically with the missing rate; rather, the presence of missing attributes triggers deterioration, whose severity depends on how current methods address incompleteness.  
For instance, FMP~\cite{jiang2024chasing} and FairSIN~\cite{yang2024fairsin} rely on attribute imputation, while FairGT~\cite{luo2024fairgt} assumes complete sensitive attributes. 
Both strategies leave model outputs correlated with sensitive information, creating privacy risks, as prior work has shown that such correlations can be exploited to reconstruct private attributes~\cite{fredrikson2015model, bai2024backdoor}. 
Therefore, developing fairness-aware GTs that ensure fairness without depending on attribute generation or complete attribute disclosure remains an urgent challenge.

\par A key challenge for GTs is to effectively encode both node and structural information, which can be addressed through multi-hop aggregation. 
This study applies spectral theory to analyse algorithmic fairness under incomplete attributes. 
Fairness is quantified by the similarity between the learned representation and the original sensitive attribute distribution~\cite{luo2024fugnn}. 
Our analysis shows that leveraging the largest eigenvalues and corresponding eigenvectors effectively encodes node and structural information, thereby capturing multi-hop neighbourhood signals. 
Spectral truncation also preserves similarity to the original sensitive attribute distribution, even with incomplete data. 
Moreover, padding missing sensitive attributes with zeros maintains consistency with the complete-data scenario while avoiding attribute reconstruction.

\par Building on this analysis, we propose \textbf{FairGE} (\textbf{Fair} \textbf{G}raph \textbf{E}ncod-ing), a fairness-aware approach for GTs in incomplete social networks.
FairGE is built on two core principles: (i) preserving fairness by selecting principal spectral components, and (ii) maintaining the independence of sensitive attributes through zero-padding. 
Concretely, FairGE first handles incomplete attributes by padding them with zeros, thereby approximating the original attribute distribution during encoding. 
It then computes the $m$ largest eigenvalues and their corresponding eigenvectors of the adjacency matrix, where spectral truncation captures multi-hop interactions while reducing the influence of non-principal components. 
The resulting encoding is integrated into GTs to produce fair node representations without reconstructing sensitive attributes.

\par The contributions are summarised as follows:

\begin{itemize}[leftmargin=0.5cm]
    \item We propose FairGE, a novel fairness-aware framework for GTs on incomplete social networks. 
    By avoiding attribute generation, FairGE mitigates privacy risks and ethical concerns linked to sensitive information restoration.
    \item We establish a theoretical foundation showing that FairGE encodes both node attributes and structural information under incompleteness. 
    Through spectral truncation and zero-padding, it ensures fairness without reconstructing sensitive data.
    \item Extensive experiments on seven real-world social network data-sets show that FairGE consistently improves fairness, yielding over $16$\% gains in both statistical parity and equality of opportunity compared with state-of-the-art methods.
\end{itemize}

\section{Related Work}
\subsection{Graph Transformers}
\par GTs have demonstrated strong potential in graph representation learning by providing global receptive fields~\cite{shehzad2024graph,lin2024rvit, zhang2024logical}. 
Models such as GraphTrans~\cite{wu2021representing} and SAN~\cite{kreuzer2021rethinking} strengthen long-range dependency modelling through attention mechanisms and spectral encodings. 
NAGphormer~\cite{chen2023nagphormer} extends this paradigm by treating nodes as sequences of token vectors and leveraging selected Laplacian eigenvectors to encode structural topology and enable multi-hop message aggregation. 
These advances have made Transformer-based graph models central to social network applications requiring expressive, non-local reasoning.

\par Despite these advances, fairness remains insufficiently addressed in GTs. 
Most existing models either overlook or amplify biases embedded in graph data due to absence of explicit fairness-aware mechanisms. 
Although FairGT~\cite{luo2024fairgt} marks an initial step by leveraging high-eigenvalue spectral attributes to promote equitable representations, its reliance on complete access to sensitive attributes limits practical deployment.
In social networks, where sensitive information (e.g., demographic attributes or regional identifiers) is often noisy, incomplete, or unavailable, this limitation becomes particularly restrictive.
This underscores the need for fairness-aware GTs that are robust to incomplete sensitive attributes, enabling equitable representation learning in real-world social network applications.

\subsection{Fairness-aware GNNs in Incomplete Social Networks}
\par Fairness-aware GNNs have attracted growing attention, with methods designed to protect or regulate sensitive attributes during learning~\cite{dong2022edits}. 
Representative models like FairGB~\cite{li2024rethinking} and FUGNN~\cite{luo2024fugnn} employ group balancing, counterfactual mixup, and spectral regularisation to jointly optimise fairness and utility. 
However, most of these approaches assume full access to sensitive attributes, which is often unrealistic in social network applications where such information is incomplete due to privacy constraints~\cite{dai2023a, chen2024fairgap}.

\par To address this challenge, recent approaches such as FairGNN~\cite{dai2023learning} and FairAC~\cite{guo2023fair} integrate attribute estimation and debiasing strategies under incomplete sensitive attributes. 
However, these methods rely on generating or inferring sensitive attributes, which can reintroduce privacy risks by reconstructing deliberately withheld data. 
Furthermore, most fairness-aware designs are tightly coupled with GNN-specific operations, limiting their applicability to GTs, that encode structure differently and capture global dependencies~\cite{zhang2024endowing, xiang2025use,zhang2024learning}. 
Consequently, enabling fairness-aware GTs in incomplete social networks remains a critical yet underexplored challenge.

\section{Preliminaries}
\subsection{Notations}
\par Unless otherwise specified, the following notation is used throughout this study: sets are denoted by copperplate uppercase letters (e.g., $\mathcal{A}$), matrices by bold uppercase letters (e.g., $\mathbf{A}$), and vectors by bold lowercase letters (e.g., $\mathbf{x}$). 
The social network is denoted as a graph, which is represented as $\mathcal{G} = (\mathcal{V}, \mathbf{A}, \mathbf{H})$, where $\mathcal{V}$ is the set of $n$ nodes, $\mathbf{A} \in \mathbb{R}^{n \times n}$ is the adjacency matrix, and $\mathbf{H} \in \mathbb{R}^{n \times d}$ is the node attribute matrix with $d$ denoting attribute dimension.

\par The spectrum of $\mathbf{A}$ reflects the graph's structural topology~\cite{bo2023a}.
Since $\mathbf{A}$ is symmetric, its eigendecomposition can be written as $\textbf{A} = \mathbf{P} \Lambda \mathbf{P}^\top$, where $\mathbf{P} = (\mathbf{p}_1, \mathbf{p}_2, \dots, \mathbf{p}_n)$ with $\mathbf{p}_i \in \mathbb{R}^{n*1}$.
The diagonal matrix $\Lambda = diag(\lambda_1, \lambda_2, ......, \lambda_n)$ contains the eigenvalues of $\textbf{A}$, ordered such that $|\lambda_1| > |\lambda_2| \geq \dots \geq |\lambda_n|$.
Here, $\mathbf{p}_i$ corresponds to $\lambda_i$, where $\mathbf{p}_1$ is associated with the dominant eigenvalue $|\lambda_1|$.

\par This indexing rules follow those of NumPy in Python.
For a matrix $\mathbf{A}$, $\mathbf{A}[i,j]$ denotes the entry at the $i$-th row and the $j$-th column, while $\mathbf{A}[i,:]$ and $\mathbf{A}[:,j]$ denote the $i$-th row and $j$-th column, respectively.
In social networks, many attributes are typically withheld by users and are therefore regarded as sensitive attributes~\cite{mehrabi2021a}.
For the attribute matrix $\mathbf{H}$, each column corresponds to one sensitive attribute, denoted by $\mathbf{H}[:,s]$.
For binary sensitive attributes, the entries take values $\mathbf{H}[i,s] \in \{0, 1\}$.
When sensitive attributes are incomplete, they denote as $\mathbf{H}'[:,s]$, and $\mathbf{H}'[:,s](0)$ denotes padding zeros for incomplete sensitive attributes.

\subsection{Fairness Evaluation Metrics}
\label{Metrics}
\par Fairness in binary classification is considered with predicted labels $\hat{y} \in \{0,1\}$, ground truth labels $y \in \{0,1\}$, and a binary sensitive attribute $s \in \{0,1\}$.
Both metrics are commonly used to assess fairness on the test set.

\noindent\textbf{Statistical Parity}~\cite{dwork2012fairness} measures the difference in positive prediction rates between sensitive groups:
\begin{equation}
\Delta_{\text{SP}} = |\mathbb{P}(\hat{y}=1 \mid \mathbf{H}[i:s]=0) - \mathbb{P}(\hat{y}=1 \mid \mathbf{H}[i:s]=1)|.
\end{equation}

\noindent\textbf{Equal Opportunity}~\cite{hardt2016equality} evaluates the gap in true positive rates across sensitive groups:
\begin{equation}
\begin{aligned}
\Delta_{\text{EO}} &= |\mathbb{P}(\hat{y}=1 \mid y=1, \mathbf{H}[i:s]=0) 
\\ &\quad\quad - \mathbb{P}(\hat{y}=1 \mid y=1, \mathbf{H}[i:s]=1)|.
\end{aligned}
\end{equation}

\subsection{Transformer} 
\par The Transformer architecture comprises multiple layers, each with a self-attention module and a position-wise feed-forward network (FFN). 
Let $\mathbf{X} = [h_1^\top, \dots, h_i^\top]^\top \in \mathbb{R}^{i \times d_m}$ denote the input to the self-attention module, where $d_m$ is the hidden dimension, and $h_j \in \mathbb{R}^{1 \times d_m}$ is the hidden representation at position $j$.
The input matrix $\mathbf{X}$ is projected using three weight matrices: $W_Q \in \mathbb{R}^{d_m \times d_K}$ (query), $W_K \in \mathbb{R}^{d_m \times d_K}$ (key), and $W_V \in \mathbb{R}^{d_m \times d_V}$ (value). 
The self-attention mechanism is computed as follows: 
\begin{equation}
    \begin{aligned}
        &Q = \mathbf{X}W_Q, \quad K = \mathbf{X}W_K, \quad V = \mathbf{X}W_V,\\
        &\text{Attn}(\mathbf{X}) = \text{Softmax}\left(\frac{QK^\top}{\sqrt{d_K}}\right)V.
    \end{aligned} 
\end{equation}

\par For simplicity, it is assumed $d_K = d_V$, focusing on the self-attention module. 
Extending this to multi-head attention is straightforward, and bias terms are omitted for clarity.

\begin{figure*}[t]
    \centering
    \includegraphics[width=0.9\textwidth]{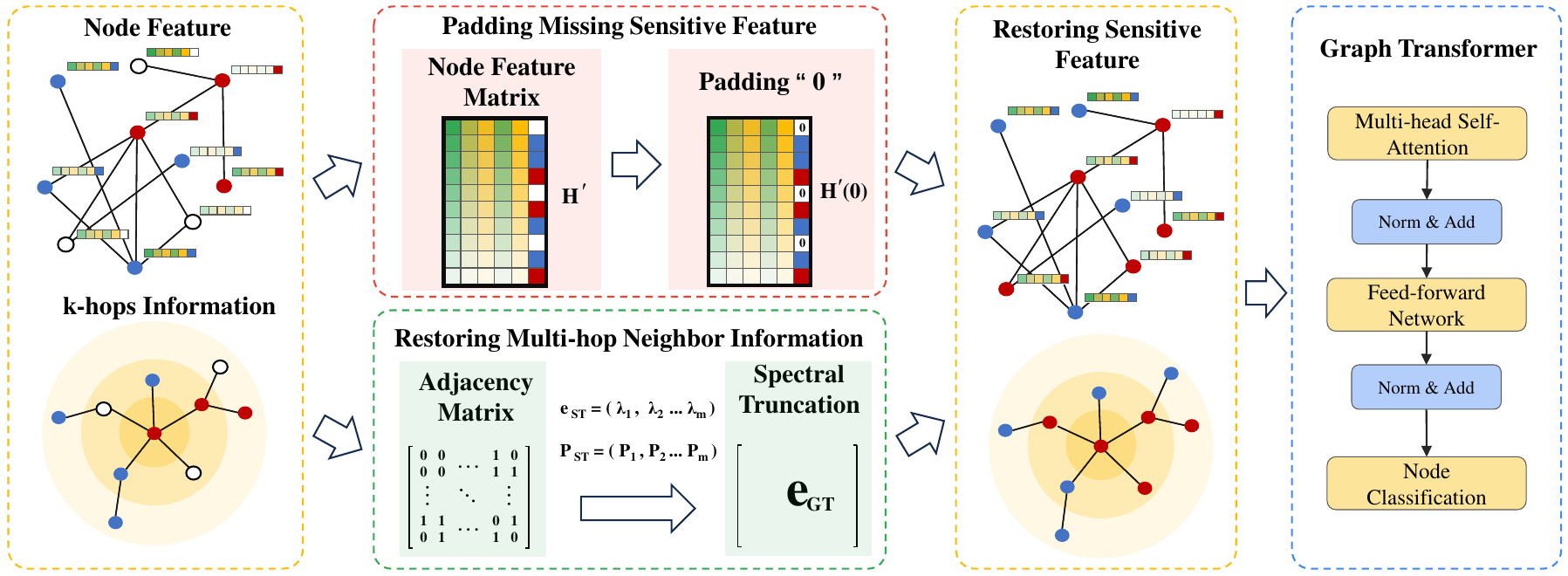}
    \caption{The illustration of FairGE. For graphs containing two sensitive attributes (represented as red and blue) alongside incomplete sensitive attributes (white), FairGE effectively restores and preserves the original sensitive attributes while also capturing multi-hop neighbour information.}
    \label{fig:illustation}
\end{figure*}

\section{Theoretical Discovery}
\par Algorithmic fairness is quantified by measuring the similarity between the inferred and the original distributions of sensitive attributes, which aligns with the principle of sensitive attribute independence.
A graph representation that preserves this similarity, even under incomplete attributes and without resorting to feature generation, can thus promote fairness in scenarios where sensitive attributes are missing.

\par We analyse the effect of structural encodings, particularly the adjacency matrix and its spectral truncation, on fairness preservation. 
Our results show that using only principal eigenvectors preserves the independence of sensitive attributes, even in incomplete social networks. 
Moreover, combining zero-padding with spectral truncation restores the original sensitive attribute distribution without relying on synthetic data, ensuring fairness in graph representation. 
Detailed proofs are provided in Appendix~\ref{Proofs}.

\subsection{Spectral Truncation as Structural Topology Encoding}
For complete sensitive attributes, prior work has demonstrated that the principal eigenvectors significantly influence the similarity between the distribution of original sensitive attributes and that of their multi-hop propagations:

\textbf{Lemma 1}~\cite{luo2024fairgt}: \textit{The similarity between the distribution of original sensitive attributes and that of multi-hop sensitive attributes exhibits a strong correlation with the principle eigenvectors.}
    \begin{equation} \label{equ:lem_1}
        \lim_{k \to \infty} \cos(\langle \mathbf{A}^k \mathbf{H}[:,s], \mathbf{H}[:,s] \rangle) = \cos(\langle \mathbf{p}_1, \mathbf{H}[:,s] \rangle). 
    \end{equation}

Building upon this observation, this study further investigates the case with incomplete sensitive attributes $\mathbf{H}'[:,s]$.

\textbf{Theorem 1}: \textit{The effect of utilising the principal eigenvector as a structural encoding, in comparison to using multiple adjacency matrices as structural topology encoding for multi-hop sensitive attributes under incomplete sensitive attributes, can be approximated.
The following equation holds:}
    \begin{equation} \label{equ:def_1.1} 
        \lim_{k \to \infty} \cos(\langle \mathbf{A}^k \mathbf{H}'[:,s], \mathbf{H}'[:,s] \rangle) = \cos(\langle \mathbf{p}_1, \mathbf{H}'[:,s] \rangle). 
    \end{equation}

\par \textbf{Theorem 1.1}: \textit{In cases where multiple maximal eigenvalues exist, the conclusions remain analogous, with further elaboration as following}:
\begin{equation}
    \begin{aligned}
        \lim_{l \to \infty} cos(\langle\mathbf{A}^k \mathbf{H}'[:,s], \mathbf{H}'[:,s]\rangle)  
        \geq \frac{1}{\sqrt{j}}\sum \nolimits_{i=1}^j{cos(\langle \mathbf{H}'[:,s], \mathbf{p}_i\rangle)}.
    \end{aligned}
\end{equation}

\par \textbf{Theorem 1.2}: \textit{The effect of non-principal eigenvalues diminishes at an exponential rate.}

\par The presence of a large spectral gap ensures that the principal eigenvectors dominate even in shallow architectures with finite depth (e.g., $k=2$ or $3$), enabling spectral truncation to effectively encode multi-hop structural information. 
Importantly, this property guarantees that the conclusions remain valid even when GTs operate under limited depth, a setting commonly adopted in practice.
Moreover, by avoiding entanglement between incomplete sensitive attributes and structural encodings, this approach preserves algorithmic fairness while simultaneously enhancing the expressiveness of graph representations.
These insights motivate the design choice of focusing on the dominant spectrum and excluding non-principal eigenvectors for fair and informative structural encoding.

\textbf{Theorem 2}: \textit{When sensitive attributes incomplete in social networks, utilising principal eigenvectors as structural topology encoding can effectively reproduce the representation results with the adjacency matrix under complete attributes. 
The following equation holds:}
\begin{equation} 
    \lim_{k \to \infty} \cos(\langle \mathbf{A}^k \mathbf{H}[:,s], \mathbf{H}'[:,s] \rangle) = \cos(\langle \mathbf{p}_1, \mathbf{H}'[:,s] \rangle). 
\end{equation}

\par The spectral truncation ensures that the encoded graph structure remains robust even under finite-depth architectures, enabling accurate representation of multi-hop structural information.
At the same time, it preserves fairness by avoiding the generation or reconstruction of incomplete sensitive attributes, thereby mitigating privacy risks while maintaining expressive graph encodings.

\subsection{Combining Padding Zeros and Spectral Truncation}
\textbf{Theorem 3}: \textit{Padding incomplete value with zeros can effectively recover the neighbouring sensitive attribute information with incomplete sensitive attributes. 
The following equation holds:}
\begin{equation}
\label{Theorem 3}
    \begin{aligned}
        &\lim_{k \to \infty} cos(\langle\mathbf{A}^k \mathbf{H}'[:,s](0), \mathbf{H}[:,s]\rangle) \\
        &\quad\quad =\lim_{k \to \infty} cos(\langle\mathbf{A}^k \mathbf{H}[:,s], \mathbf{H}[:,s]\rangle).
    \end{aligned}
\end{equation}

\par In addition, the padding value $0$ does not represent a fixed property; rather, it aligns with our design principle of avoiding synthetic data generation.
By analogy to \textbf{Theorem 1.2}, a similar phenomenon holds for incomplete sensitive attributes. 
Even with a limited number of layers ($k =2$ or $3$), this conclusion remains valid due to the significant spectral gap in real-world datasets ($\lambda_1 >> \lambda_2$), ensuring the dominance of the principal eigenvector in the multi-hop propagation.
This theorem is valid if female is $0$ or male is $0$.
In summary, under the scenario of incomplete social networks, the sensitive attributes of multi-hop neighbour information are predominantly influenced by the eigenvector corresponding to the largest magnitude eigenvalue (or multiple eigenvectors associated with the largest magnitude eigenvalue). 
In addition in cases where sensitive attributes are incomplete, padding the incomplete attributes with zeros enhances the fairness and utility of these principal eigenvectors in the model. 
Therefore, in the context of incomplete social networks, this study proposes that combining spectral truncation with simple zero-padding significantly improves fairness.

\section{Design of FairGE}
\par FairGE for fairness-aware node classification in GTs integrates three key components. 
First, zero padding addresses incomplete sensitive attributes, mitigating privacy risks while, as shown in our theoretical analysis, preserving multi-hop neighbour information and sensitive attribute independence. 
Second, spectral truncation leverages principal eigenvectors to reconstruct multi-hop signals, improving both fairness and accuracy. 
Finally, the combined encoding from zero padding and spectral truncation is integrated into the GT, enabling efficient and fair node classification. 
Figure~\ref{fig:illustation} illustrates the model architecture.


\subsection{Padding incomplete Sensitive attributes}
\par In graph mining tasks, node attribute information is crucial. 
To address incomplete sensitive attributes, FairGE proposes a zero-padding encoding. 
This method captures multi-hop information that mirrors original sensitive attribute patterns.

\par For a given node $v$, its $k$-hop neighbourhood is denoted as $\mathcal{V}^{(k)}$, with $\mathcal{V}^{(0)} = \{v\}$ representing the node itself. 
In GTs with incomplete attributes, the $k$-hop neighbourhood information is transformed into an attribute embedding $\mathbf{H'}^{(k)}_v$ via an aggregation operator $\Phi$, which integrates information from nodes across multiple hop distances. 
Formally, the aggregation for node $v$ is defined as:
\begin{equation}
  \label{equ:k-hop}
  \mathbf{H'}^{(k)}_v=\Phi(v,\mathcal{V}^{(1)},..., \mathcal{V}^{(k)}).
\end{equation}

\par Using this formulation, FairGE computes the multi-hop attribute information for any node.
To simplify the node attribute encoding, FairGE decomposes the sequence $\mathbf{S} = (\mathbf{H'}^{(0)}, \mathbf{H'}^{(1)}, \dots, \mathbf{H'}^{(k)})$, where $\mathbf{H'}^{(k)} \in \mathbb{R}^{n \times d}$ denotes the multi-hop attribute matrix and $\mathbf{H'}^{(0)}$ corresponds to the original attribute matrix. 
Given adjacency matrix $\mathbf{A}$ of $\mathcal{G}$, the multi-hop attribute matrix is computed as:
\begin{equation}
  \mathbf{H'}^{(k)} = \mathbf{A}^k\mathbf{H'}.
\end{equation}

\par As shown in \textbf{Theorem~3} (see Equation~\eqref{Theorem 3}), the zero-padding encoding $\textbf{H}'(0)$ closely resembles the original sensitive attribute structure. 
Encoding multi-hop information using zero-padding helps capture the correlations across hops and contributes significantly to fairness in GTs.

\subsection{Restoring Multi-hop Neighbour Information}
\par Beyond node attribute encoding, structural topology encoding is also crucial for capturing valuable graph information. 
To restore multi-hop neighbour information, FairGE introduces a novel structural topology encoding, which utilises spectral truncation to process multi-hop information.
Spectral graph theory reveals that algebraic connectivity (the second smallest eigenvalue) and spectral radius (the largest eigenvalue) are deeply connected to the geometric properties of the graph~\cite{kreuzer2021rethinking}. 

\par Drawing from \textbf{Theorem 1} and \textbf{Theorem 2}, FairGE adopts spectral truncation (ST) of the adjacency matrix to recover the incomplete sensitive information of multi-hop neighbours.
Specifically, FairGE selects the eigenvectors $\mathbf{P}_\text{ST}$ corresponding to the $m$ largest eigenvalues $\mathbf{e}_\text{ST}$, as follows:
\begin{equation}
  \begin{aligned}
  \mathbf{e}_\text{ST} = (\lambda_1, \lambda_2, ...... ,\lambda_m),\quad\mathbf{P}_\text{ST} = (\mathbf{p_1}, \mathbf{p_2}, ...... , \mathbf{p_m}).
  \end{aligned}
  \label{equ: structural information}
\end{equation}

\par In FairGE, the hyperparameter $m$ dictates the number of principal eigenvectors retained from the graph spectrum. 
To compute these efficiently, FairGE uses the Arnoldi Package~\cite{cai2021a} for its high computational efficiency and accuracy.
While full eigendecomposition typically has a cubic time complexity $O(n^3)$, this package reduces FairGE's  complexity to $O(nm^2)$ in time and $O(n)$ in space.
Even with small values of $m$, FairGE remains highly effective with minimal performance fluctuations.

\subsection{FairGE for Node Classification}
\par In the FairGE framework, the process begins by concatenating a matrix derived from the adjacency matrix's eigendecomposition to encode the structural topology, as defined by Equation~\eqref{equ: structural information}. 
To address incomplete sensitive attributes, FairGE then applies zero padding, denoted as $\mathbf{H}'(0)$.

\par Next, FairGE applies a position encoding to the eigenvalues using the following sinusoidal and cosine functions:  
\begin{equation}
    \begin{aligned}
        &\rho(\lambda_{2i}) = sin(\lambda_{2i}/10000^{2i/d}), \\
        &\rho(\lambda_{2i+1}) = cos(\lambda_{2i+1}/10000^{2i/d}),
    \end{aligned}
\label{equ:encoding_function}
\end{equation}
resulting in a position-encoded (PE) eigenvalue sequence: 
\begin{equation}
    \mathbf{e}_\text{PE} = (\rho(\lambda_1), \rho(\lambda_2), ......, \rho(\lambda_m)).
\end{equation}

\par FairGE then projects this transformed sequence, which encodes the structural topology, into the Transformer architecture \cite{vaswani2017attention}. 
The Transformer consists of multi-head self-attention (MHA) and a position-wise feed-forward network (FFN), with Layer Normalization (LN) applied before each block. 
The FFN is composed of two linear layers with a GELU activation function:
\begin{equation}
    \begin{aligned}
        &\mathbf{e}_\text{MHA}= \textbf{MHA}(\textbf{LN}(\mathbf{e}_\text{PE}))+\mathbf{e}_\text{PE}, \\
        &\mathbf{e}_\text{GT}= \textbf{FFN}(\textbf{LN}(\mathbf{e}_\text{MHA}))+ \mathbf{e}_\text{MHA}.
    \end{aligned}
\label{equ:Transformer}
\end{equation}
\begin{figure*}[t]
	\centering
    \subfigure {
		\begin{minipage}[b]{0.28\textwidth}
			\centering
			\includegraphics[width=1.0\textwidth]{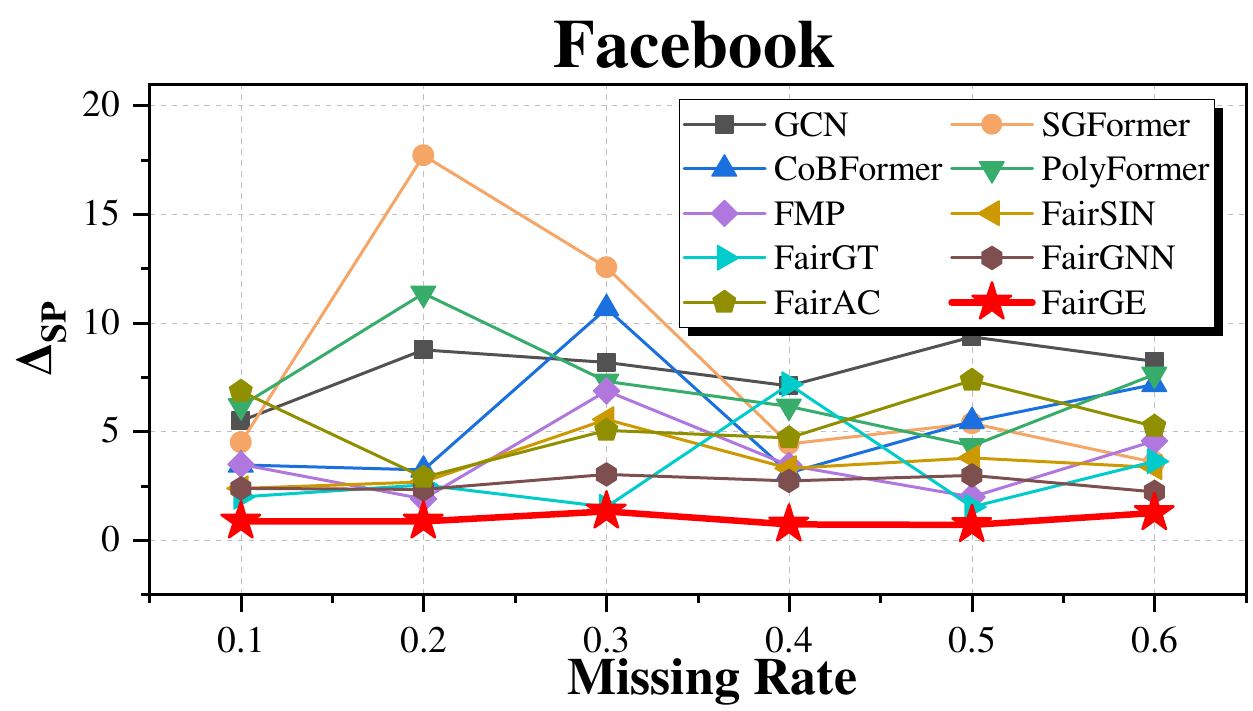}
		\end{minipage}
	}
    \subfigure {
		\begin{minipage}[b]{0.28\textwidth}
			\centering
			\includegraphics[width=1.0\textwidth]{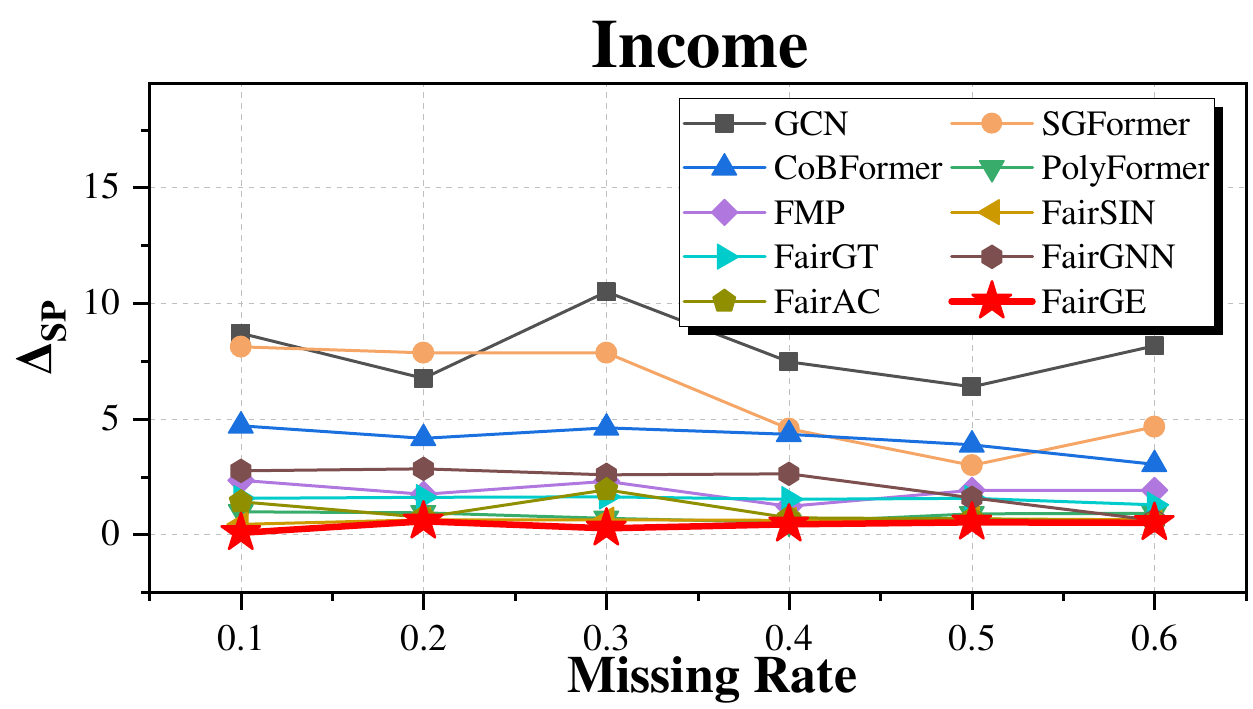}
		\end{minipage}
	}
	\subfigure {
		\begin{minipage}[b]{0.28\textwidth}
			\centering
			\includegraphics[width=1.0\textwidth]{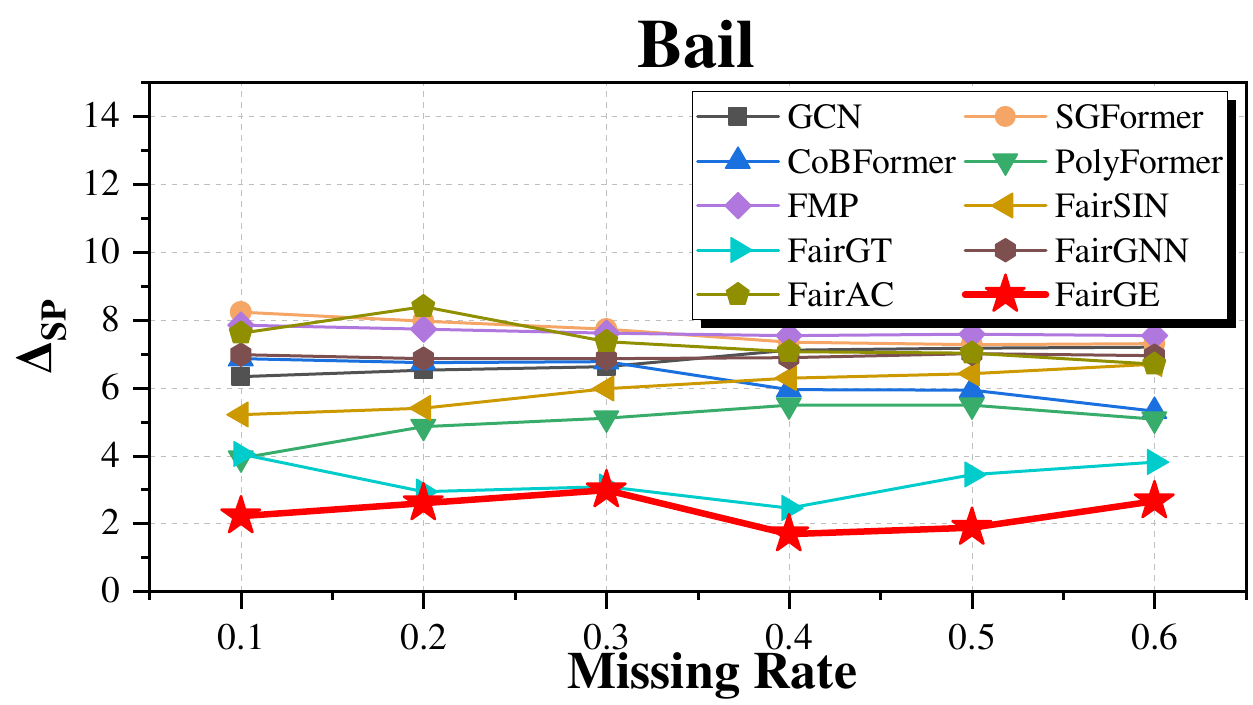}
		\end{minipage}
	}
    \subfigure {
		\begin{minipage}[b]{0.28\textwidth}
			\centering
			\includegraphics[width=1.0\textwidth]{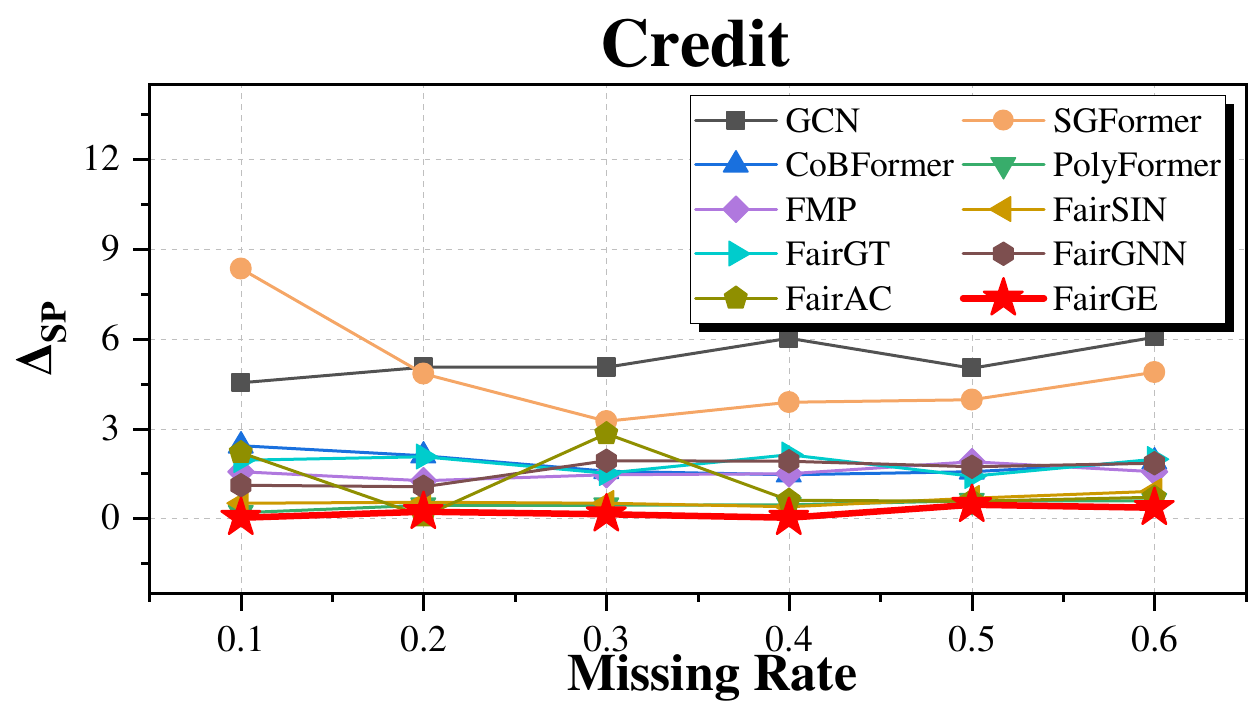}
		\end{minipage}
	}
	\subfigure {
		\begin{minipage}[b]{0.28\textwidth}
			\centering
			\includegraphics[width=1.0\textwidth]{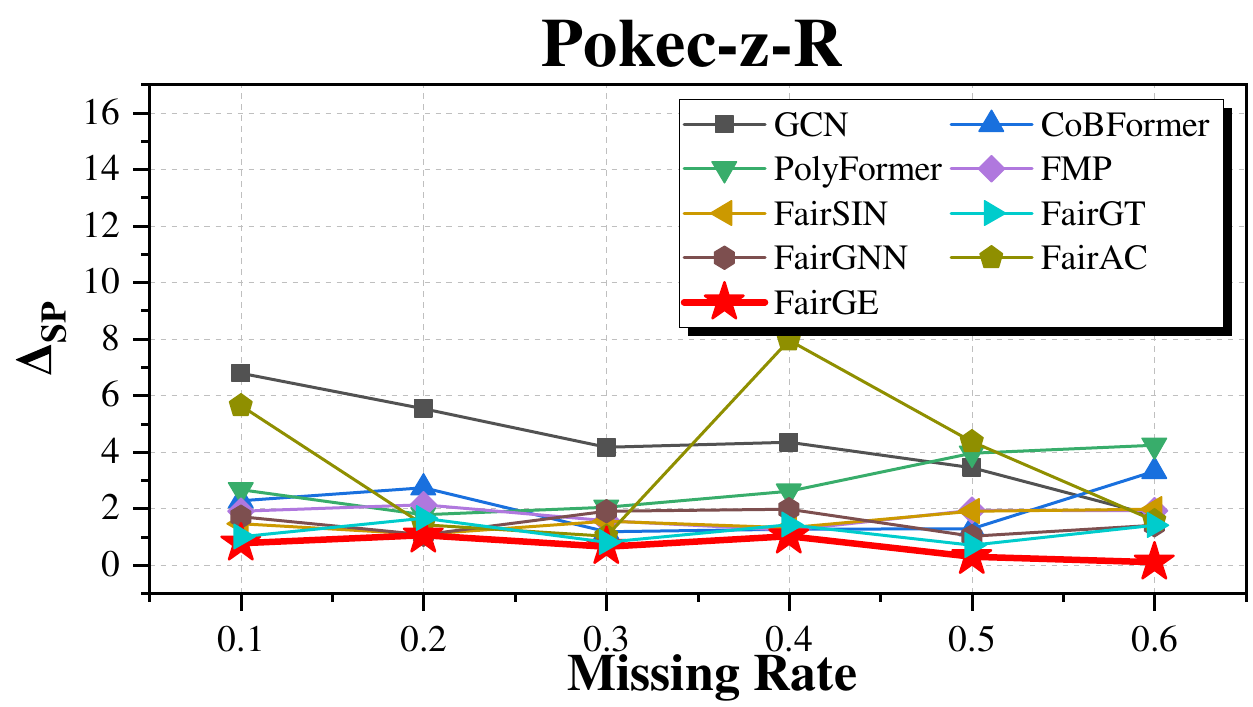}
		\end{minipage}
	} 
	\subfigure {
		\begin{minipage}[b]{0.28\textwidth}
			\centering
			\includegraphics[width=1.0\textwidth]{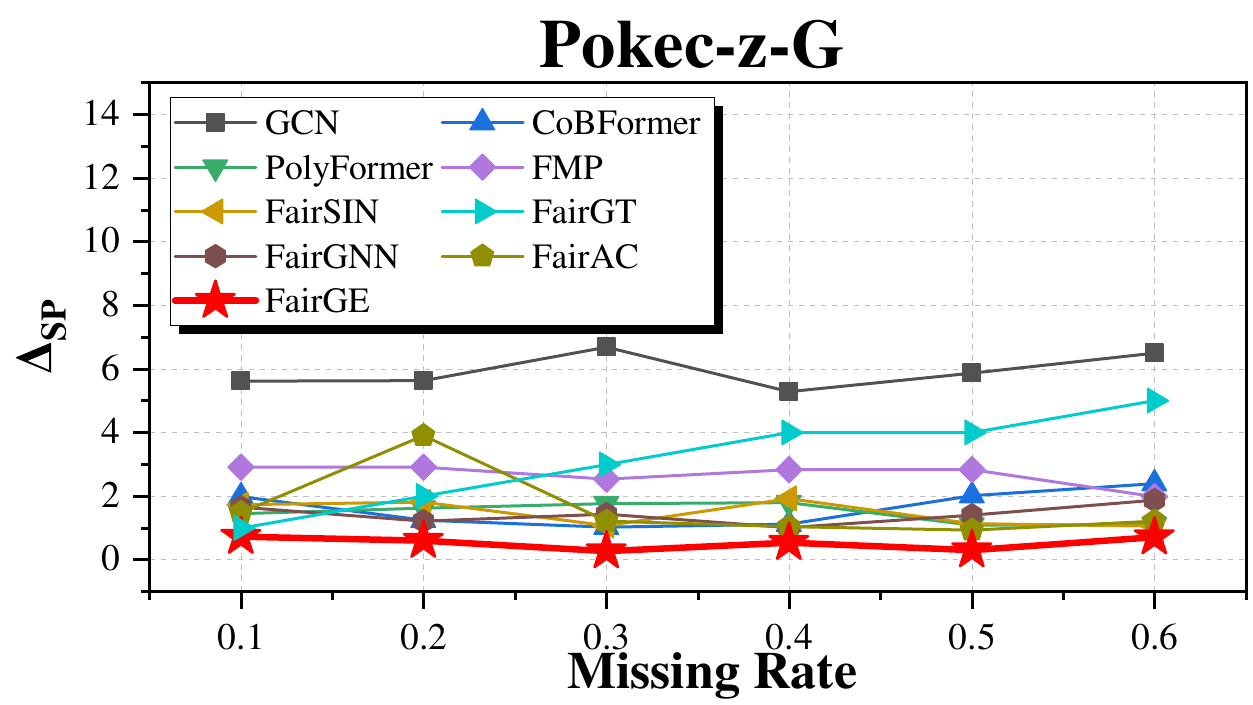}
		\end{minipage}
	}
    \subfigure {
		\begin{minipage}[b]{0.28\textwidth}
			\centering
			\includegraphics[width=1.0\textwidth]{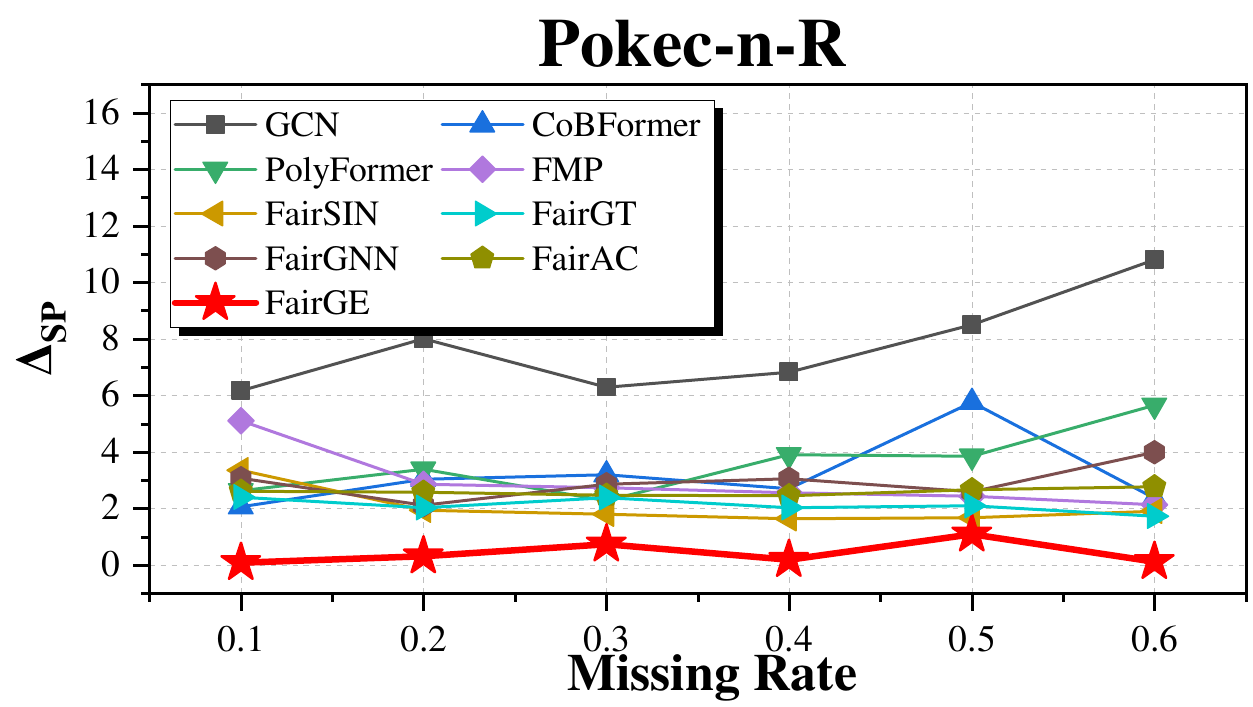}
		\end{minipage}
	}
    \subfigure {
		\begin{minipage}[b]{0.28\textwidth}
			\centering
			\includegraphics[width=1.0\textwidth]{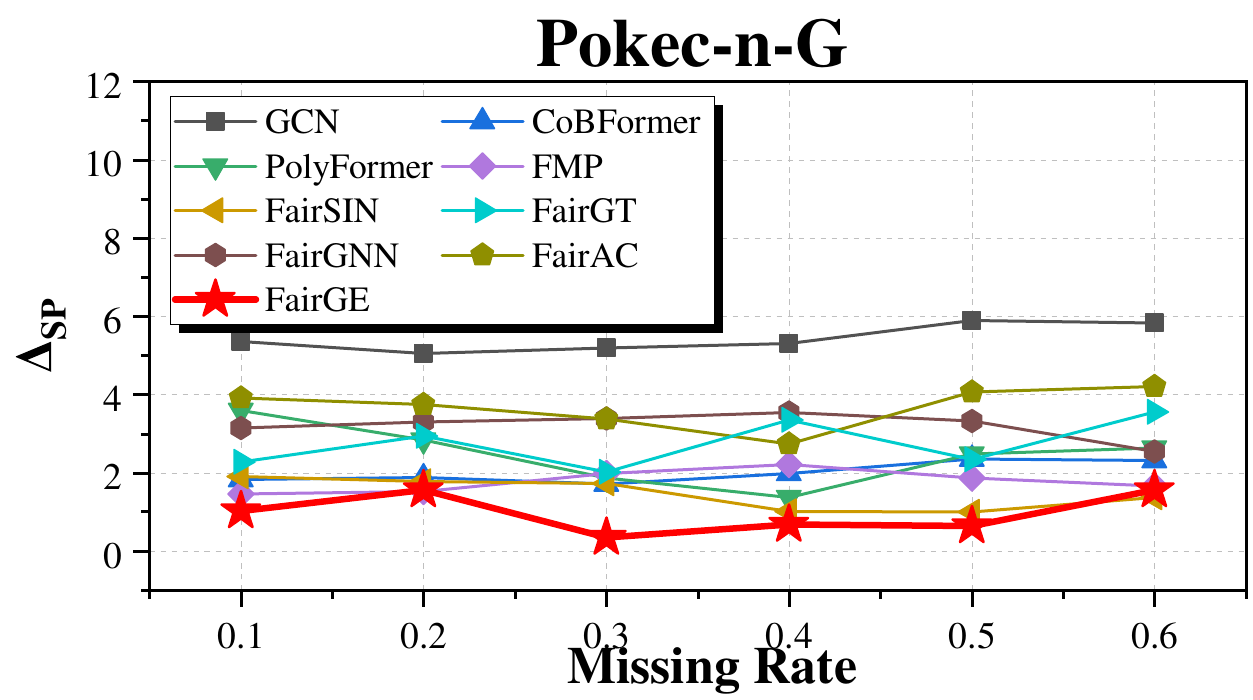}
		\end{minipage}
	}
    \subfigure {
		\begin{minipage}[b]{0.28\textwidth}
			\centering
			\includegraphics[width=1\textwidth]{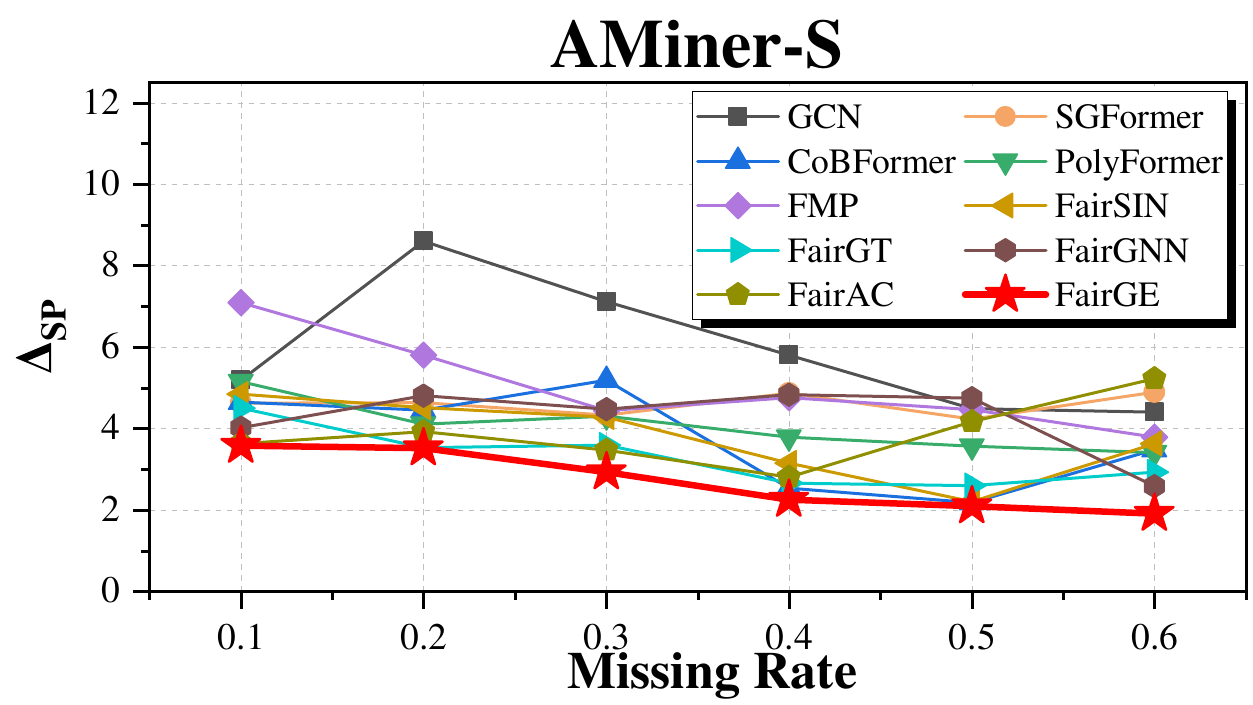}
		\end{minipage}
	} 
  \caption{Comparison of $\Delta_{\text{SP}}$ between FairGE and baseline models.}
  \label{fig:results_sp}
\end{figure*}

\begin{figure*}[t]
	\centering
	\subfigure {
		\begin{minipage}[b]{0.28\textwidth}
			\centering
			\includegraphics[width=1.0\textwidth]{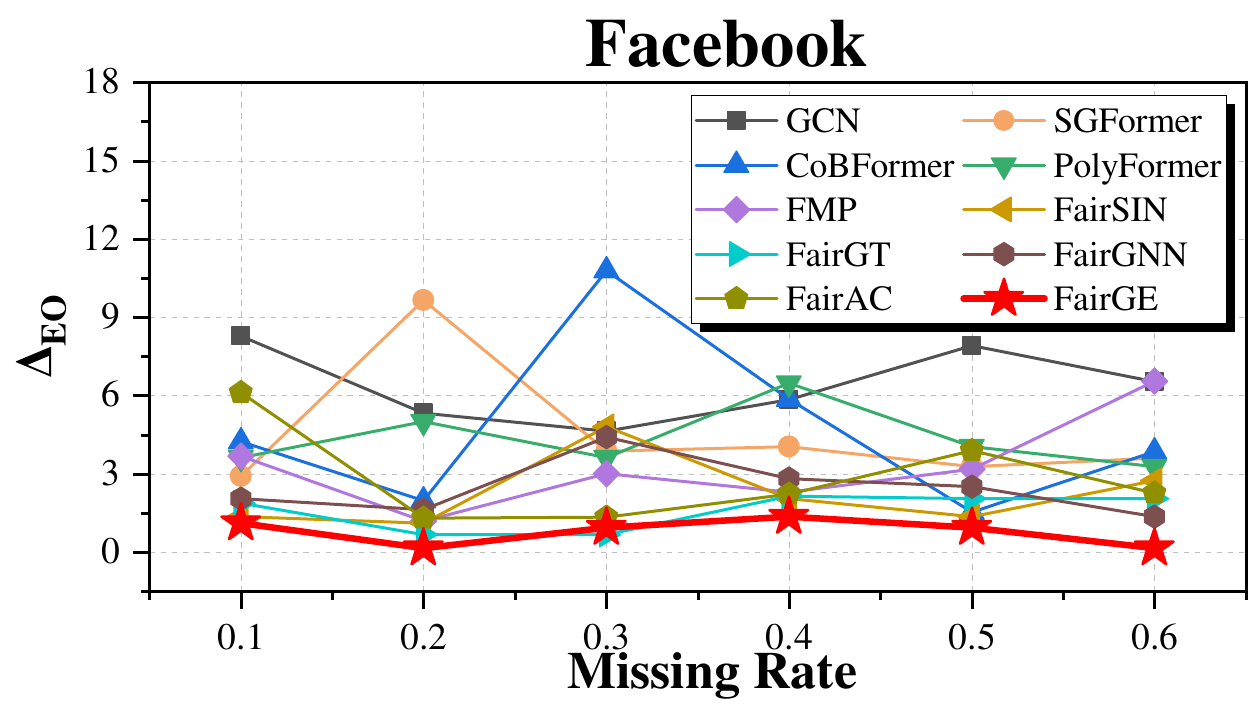}
		\end{minipage}
	}
    \subfigure {
		\begin{minipage}[b]{0.28\textwidth}
			\centering
			\includegraphics[width=1.0\textwidth]{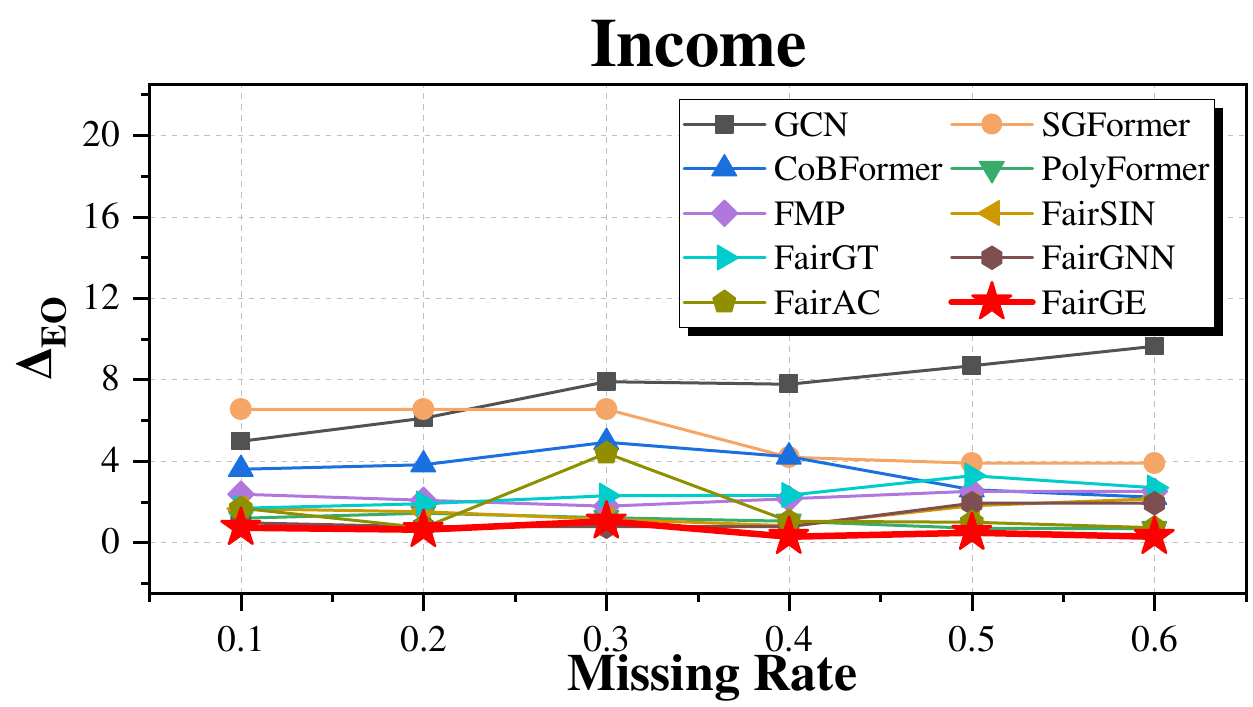}
		\end{minipage}
	}
	\subfigure {
		\begin{minipage}[b]{0.28\textwidth}
			\centering
			\includegraphics[width=1.0\textwidth]{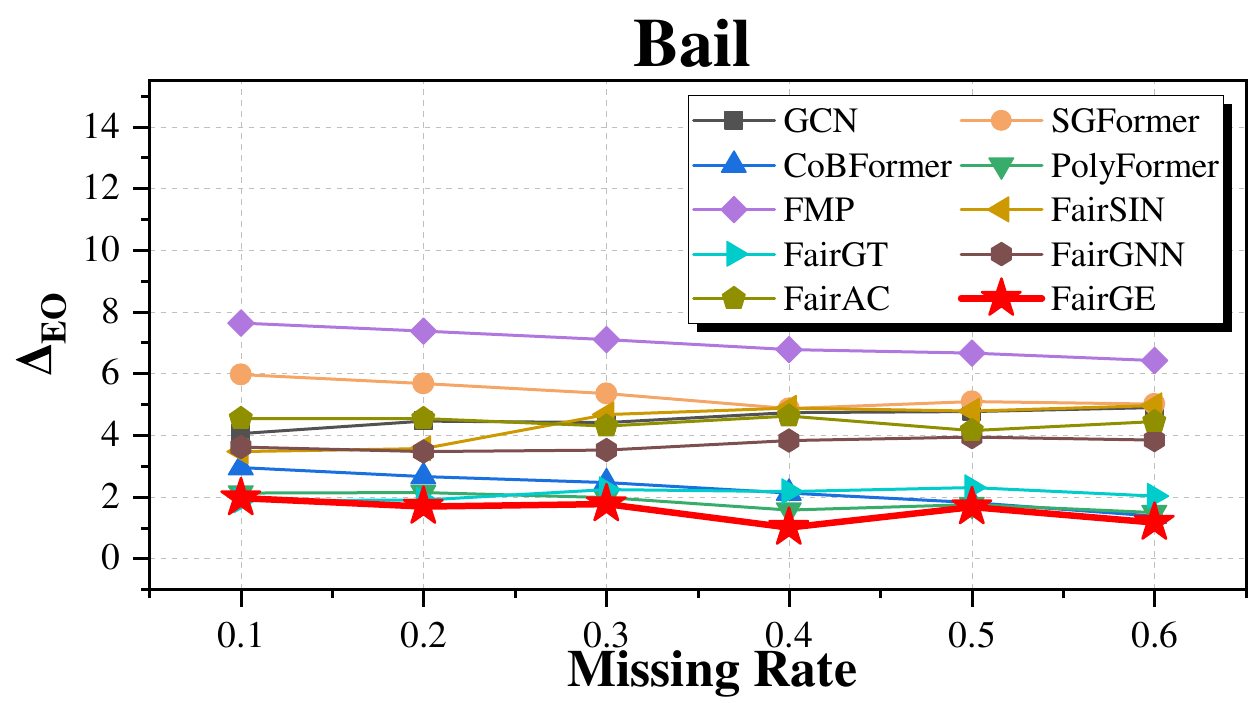}
		\end{minipage}
	}
    \subfigure {
		\begin{minipage}[b]{0.28\textwidth}
			\centering
			\includegraphics[width=1.0\textwidth]{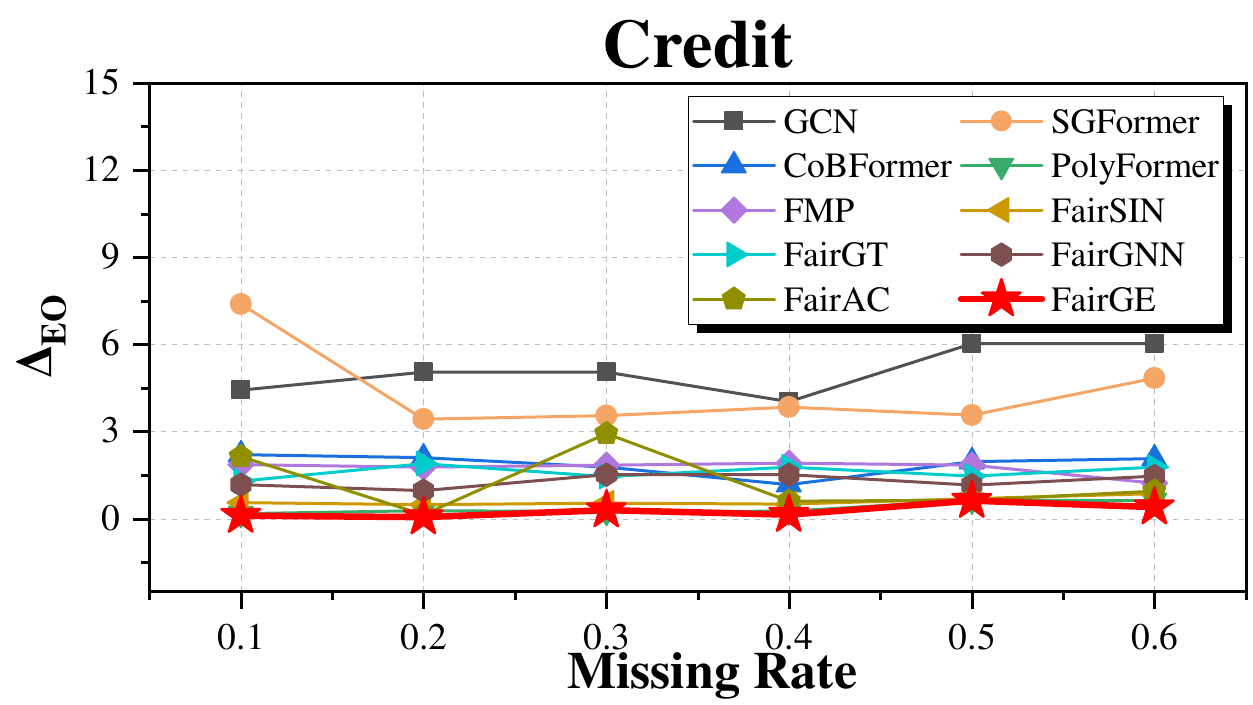}
		\end{minipage}
	}
	\subfigure {
		\begin{minipage}[b]{0.28\textwidth}
			\centering
			\includegraphics[width=1.0\textwidth]{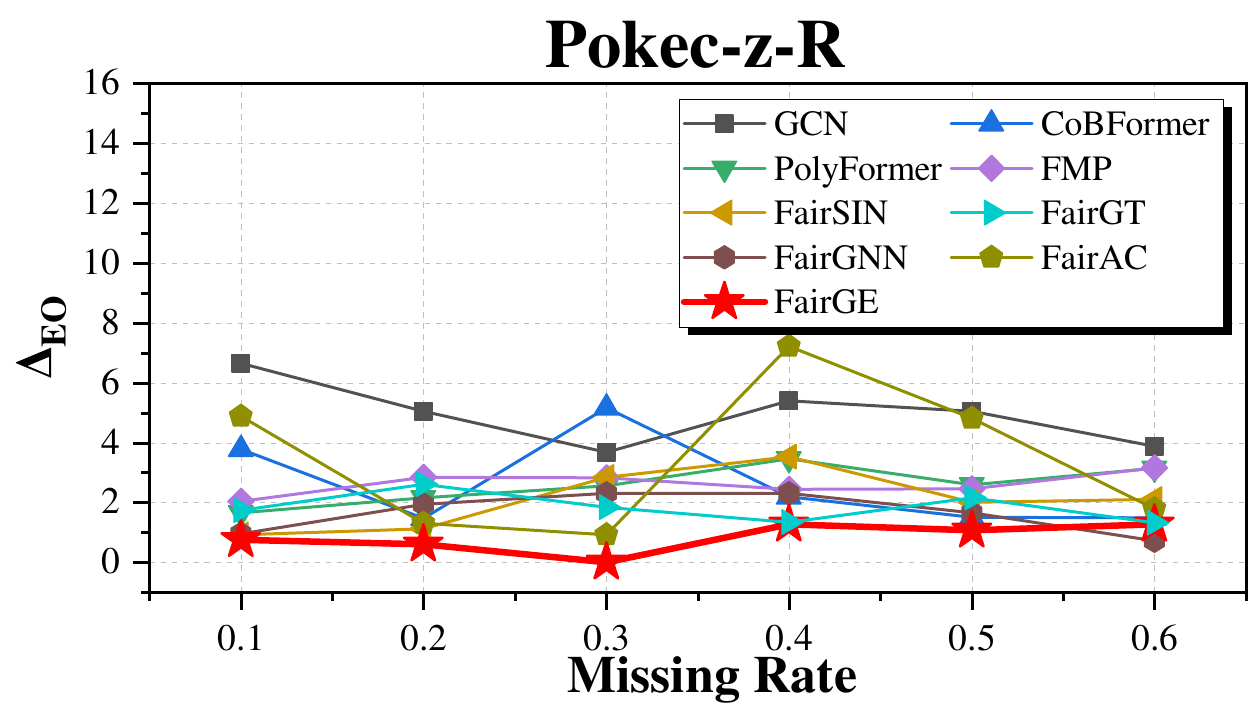}
		\end{minipage}
	} 
	\subfigure {
		\begin{minipage}[b]{0.28\textwidth}
			\centering
			\includegraphics[width=1.0\textwidth]{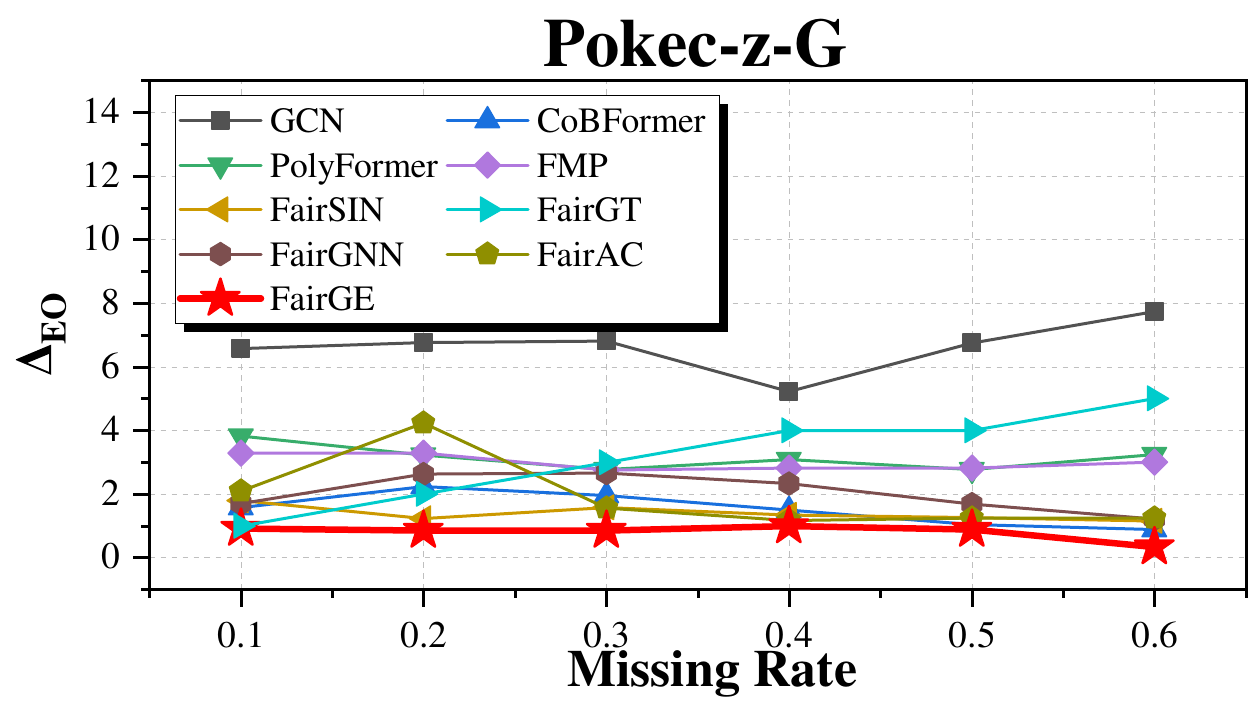}
		\end{minipage}
	}
    \subfigure {
		\begin{minipage}[b]{0.28\textwidth}
			\centering
			\includegraphics[width=1.0\textwidth]{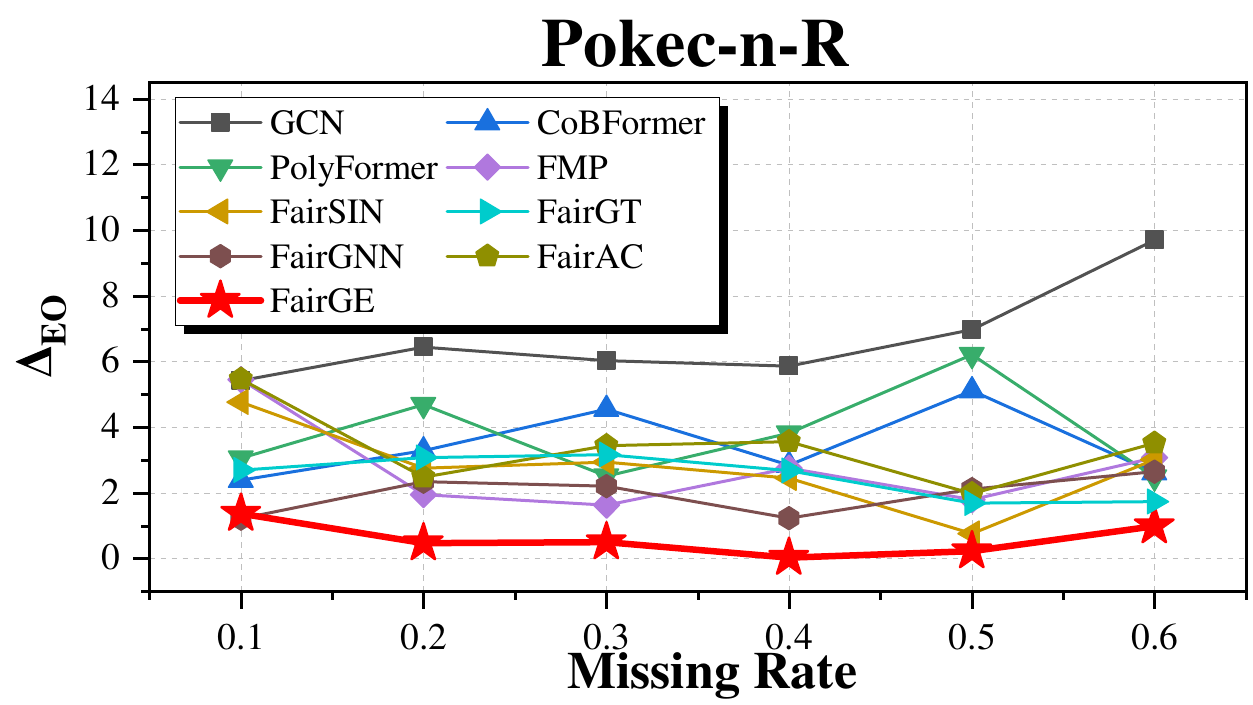}
		\end{minipage}
	}
    \subfigure {
		\begin{minipage}[b]{0.28\textwidth}
			\centering
			\includegraphics[width=1.0\textwidth]{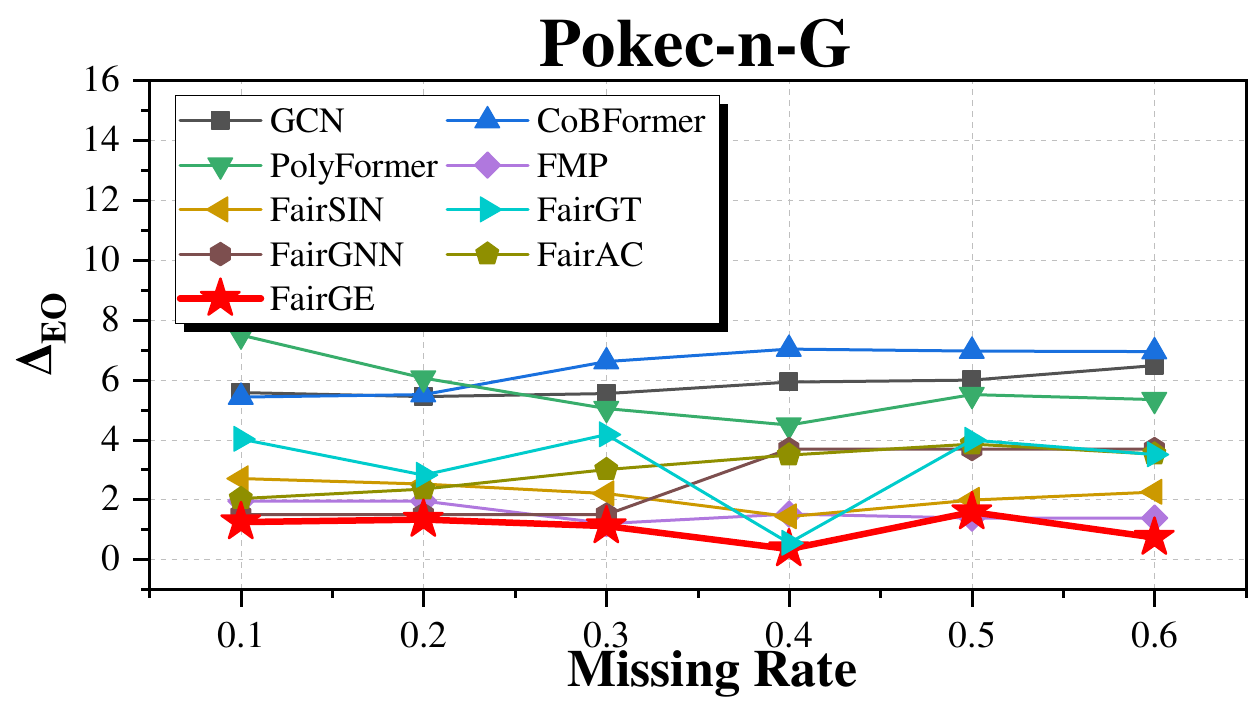}
		\end{minipage}
	}
    \subfigure {
		\begin{minipage}[b]{0.28\textwidth}
			\centering
			\includegraphics[width=1\textwidth]{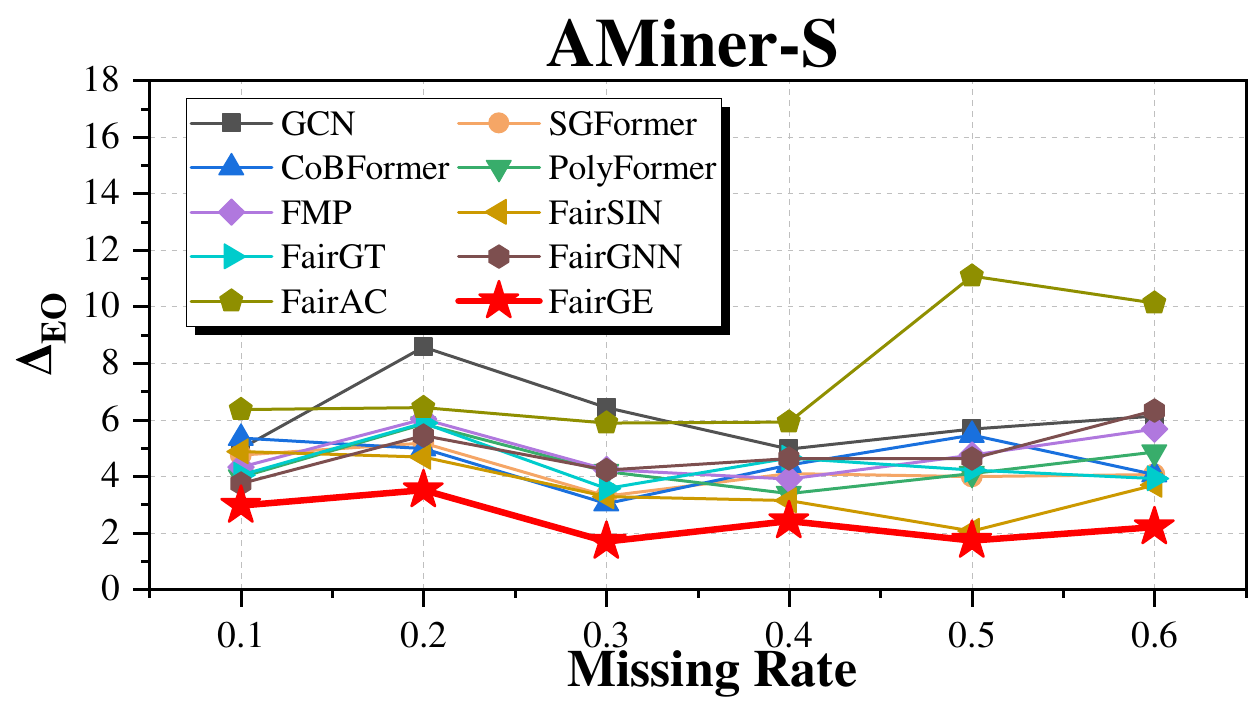}
		\end{minipage}
	} 
  \caption{Comparison of $\Delta_{\text{EO}}$ between FairGE and baseline models.}
  \label{fig:results_eo}
\end{figure*}

\par Finally, FairGE refines the spectral representation using the learned embedding $\mathbf{e}_\text{GT}$ and transforms the zero-padded attributes $\mathbf{H}'(0)^{(l)}$ at each layer, where $\mathbf{W}^{(l-1)}$ denotes the trainable transformation matrix of the $(l-1)$-th layer and $\sigma(\cdot)$ is a non-linear activation function: 
\begin{equation}
  \begin{aligned}
        &\mathbf{H}_\text{train}^{(l)} = \mathbf{P}_\text{ST} \cdot (e_\text{GT} \odot \mathbf{P}_\text{ST}^\top \mathbf{H}'(0)^{(l)}),\\
        &\mathbf{H}^{(l)}= \sigma ((\mathbf{H}^{(l-1)} || \mathbf{H}_\text{train}^{(l-1)}) \mathbf{W}^{(l-1)}).
  \end{aligned}
\end{equation}

\par By stacking multiple GT layers, FairGE effectively learns node representations that incorporate graph information, improving the model’s performance even with incomplete sensitive attributes.

\begin{table}[t]
  \centering
  \small
  \tabcolsep=0.15cm
  \renewcommand{\arraystretch}{0.85}
  \caption{Statistics of the seven real-world datasets.}
  \begin{tabular}{lllll}
  \toprule
  \textbf{Dataset}    & \textbf{\# Nodes} & \textbf{\# Edges} & \textbf{Sensitive attribute} & \textbf{Label}\\
    \midrule
      \textbf{Facebook} & $1,045$   & $53,498$  & Gender        & Education \\
      \textbf{Income}   & $14,821$  & $100,483$ & Race          & Income \\
      \textbf{Bail}     & $18,876$  & $321,308$ & Race          & Recidivism \\
      \textbf{Credit}   & $30,000$  & $137,377$ & Age           & Default \\
      \textbf{Pokec-z}  & $67,797$  & $882,765$ & Region/Gender & Field \\
      \textbf{Pokec-n}  & $66,569$  & $729,129$ & Region/Gender & Field \\
      \textbf{AMiner-S} & $39,424$  & $52,460$  & Affiliation   & Field \\
    \bottomrule
  \end{tabular}
  \label{tab:datasets}
\end{table}

\begin{table*}[t]
\centering
\footnotesize
\caption{
Comparison results of FairGE and baseline models on the Credit and AMiner-S datasets.  
\textcolor{red}{\textbf{Red}} numbers denote the best performance, and \textcolor{blue}{\underline{blue}} numbers denote the second best.
}
\tabcolsep=0.25cm
\renewcommand{\arraystretch}{0.9}
\begin{tabular}{l|ccc|ccc|ccc}
 \toprule
 Missing Rate & \multicolumn{3}{c}{\textbf{Credit(0.1)}} & \multicolumn{3}{c}{\textbf{Credit(0.2)}} & \multicolumn{3}{c}{\textbf{Credit(0.3)}} \\
 Metrics & ACC(\%)$_{\uparrow}$ & $\Delta_\text{SP}$(\%)$_{\downarrow}$ & $\Delta_\text{EO}$(\%)$_{\downarrow}$ & ACC(\%)$_{\uparrow}$ & $\Delta_\text{SP}$(\%)$_{\downarrow}$ & $\Delta_\text{EO}$(\%)$_{\downarrow}$ & ACC(\%)$_{\uparrow}$ & $\Delta_\text{SP}$(\%)$_{\downarrow}$ & $\Delta_\text{EO}$(\%)$_{\downarrow}$\\
 \midrule
 GCN & $75.70\pm0.82$ & $4.54\pm0.31$ & $4.44\pm1.25$ & $75.37\pm0.96$ & $5.06\pm0.48$ & $5.06\pm0.22$ & $75.37\pm1.34$ & $5.06\pm0.73$ & $5.06\pm0.29$ \\
 SGFormer & $72.23\pm1.33$ & $8.35\pm0.27$ & $7.39\pm0.69$ & $72.70\pm0.84$ & $4.85\pm0.54$ & $3.43\pm0.19$ & $73.23\pm0.91$ & $3.26\pm0.46$ & $3.55\pm0.16$ \\
 CoBFormer & $75.67\pm0.95$ & $2.44\pm0.39$ & $2.22\pm1.18$ & $75.39\pm0.73$ & $2.11\pm0.49$ & $2.11\pm0.25$ & $76.01\pm1.02$ & $1.59\pm0.59$ & $1.78\pm0.24$ \\
 Polynormer & $77.24\pm1.08$ & \textcolor{blue}{\underline{$0.20\pm0.13$}} & \textcolor{blue}{\underline{$0.19\pm0.12$}} & $76.99\pm0.58$ & $0.46\pm0.37$ & $0.28\pm0.26$ & $76.21\pm1.17$ & \textcolor{blue}{\underline{$0.45\pm0.43$}} & \textcolor{red}{$\mathbf{0.26\pm0.19}$} \\
 FMP & $75.96\pm1.21$ & $1.57\pm0.42$ & $1.87\pm1.36$ & $76.00\pm0.65$ & $1.28\pm0.50$ & $1.78\pm0.33$ & $76.11\pm1.29$ & $1.47\pm0.61$ & $1.85\pm0.28$ \\
 FairSIN & \textcolor{blue}{\underline{$77.64\pm1.14$}} & $0.53\pm0.23$ & $0.57\pm0.77$ & $77.72\pm0.68$ & $0.56\pm0.46$ & $0.49\pm0.21$ & $77.50\pm1.32$ & $0.52\pm0.50$ & $0.54\pm0.32$ \\
 FairGT & $77.27\pm1.47$ & $1.96\pm0.32$ & $1.30\pm1.03$ & $77.26\pm0.79$ & $2.08\pm0.45$ & $1.91\pm0.24$ & $77.29\pm1.15$ & $1.52\pm0.48$ & $1.46\pm0.26$ \\
 FairGNN & $75.06\pm0.78$ & $1.12\pm0.18$ & $1.18\pm1.15$ & $74.29\pm0.64$ & $1.07\pm0.59$ & $0.98\pm0.36$ & $75.10\pm0.92$ & $1.94\pm0.71$ & $1.53\pm0.33$ \\
 FairAC & $77.18\pm1.03$ & $2.21\pm0.26$ & $2.15\pm0.87$ & \textcolor{blue}{\underline{$77.87\pm0.76$}} & \textcolor{red}{$\mathbf{0.09\pm0.05}$} & \textcolor{blue}{\underline{$0.15\pm0.12$}} & \textcolor{red}{$\mathbf{77.67\pm1.08}$} & $2.84\pm0.54$ & $2.93\pm0.28$ \\
 \textbf{FairGE} & \textcolor{red}{$\mathbf{77.72\pm0.97}$} & \textcolor{red}{$\mathbf{0.03\pm0.02}$} & \textcolor{red}{$\mathbf{0.12\pm0.10}$} & \textcolor{red}{$\mathbf{78.08\pm0.67}$} & \textcolor{blue}{\underline{$0.24\pm0.13$}} & \textcolor{red}{$\mathbf{0.07\pm0.04}$} & \textcolor{red}{$\mathbf{77.67\pm1.35}$} & \textcolor{red}{$\mathbf{0.16\pm0.12}$} & \textcolor{blue}{\underline{$0.30\pm0.09$}} \\
 \midrule
 \midrule
 Missing Rate & \multicolumn{3}{c}{\textbf{Credit(0.4)}} & \multicolumn{3}{c}{\textbf{Credit(0.5)}} & \multicolumn{3}{c}{\textbf{Credit(0.6)}} \\
 \midrule
 GCN & $75.85\pm1.45$ & $6.03\pm0.29$ & $4.04\pm1.23$ & $74.85\pm0.98$ & $5.03\pm0.47$ & $6.04\pm0.31$ & $74.85\pm1.12$ & $6.06\pm0.69$ & $6.04\pm0.24$ \\
 SGFormer & $73.40\pm0.89$ & $3.89\pm0.12$ & $3.85\pm0.56$ & $73.34\pm0.78$ & $3.98\pm0.51$ & $3.58\pm0.19$ & $72.37\pm0.87$ & $4.89\pm0.32$ & $4.85\pm0.15$ \\
 CoBFormer & $76.08\pm1.02$ & $1.47\pm0.34$ & $1.18\pm0.78$ & $75.92\pm0.67$ & $1.58\pm0.45$ & $1.97\pm0.21$ & $76.30\pm0.99$ & $1.91\pm0.53$ & $2.08\pm0.12$ \\
 Polynormer & $76.18\pm1.23$ & $0.48\pm0.34$ & \textcolor{blue}{\underline{$0.27\pm0.18$}} & $76.61\pm0.67$ & $0.61\pm0.56$ & \textcolor{blue}{\underline{$0.65\pm0.23$}} & \textcolor{blue}{\underline{$77.62\pm1.01$}} & \textcolor{blue}{\underline{$0.60\pm0.42$}} & \textcolor{blue}{\underline{$0.64\pm0.18$}} \\
 FMP & $75.59\pm0.99$ & $1.51\pm0.23$ & $1.92\pm1.45$ & $75.55\pm0.87$ & $1.91\pm0.34$ & $1.86\pm0.67$ & $75.48\pm0.56$ & $1.57\pm0.78$ & $1.23\pm0.21$ \\
 FairSIN & $77.32\pm1.11$ & \textcolor{blue}{\underline{$0.40\pm0.23$}} & $0.51\pm0.24$ & \textcolor{blue}{\underline{$77.21\pm0.69$}} & $0.69\pm0.45$ & $0.70\pm0.21$ & $77.33\pm1.23$ & $0.92\pm0.57$ & $0.88\pm0.31$ \\
 FairGT & $77.12\pm1.34$ & $2.15\pm0.32$ & $1.78\pm0.68$ & $77.09\pm0.72$ & $1.43\pm0.48$ & $1.48\pm0.24$ & $76.76\pm1.19$ & $2.00\pm0.46$ & $1.78\pm0.27$ \\
 FairGNN & $75.03\pm0.76$ & $1.92\pm0.18$ & $1.52\pm1.15$ & $76.39\pm0.64$ & $1.75\pm0.59$ & $1.16\pm0.36$ & $75.66\pm0.92$ & $1.86\pm0.71$ & $1.47\pm0.33$ \\
 FairAC & \textcolor{red}{$\mathbf{77.75\pm1.03}$} & $0.62\pm0.26$ & $0.62\pm0.87$ & \textcolor{red}{$\mathbf{77.77\pm0.76}$} & \textcolor{blue}{\underline{$0.59\pm0.44$}} & \textcolor{blue}{\underline{$0.65\pm0.22$}} & $77.56\pm1.08$ & $0.72\pm0.54$ & $0.96\pm0.28$ \\
 \textbf{FairGE} & \textcolor{blue}{\underline{$77.60\pm0.97$}} & \textcolor{red}{$\mathbf{0.04\pm0.03}$} & \textcolor{red}{$\mathbf{0.15\pm0.09}$} & $77.13\pm0.67$ & \textcolor{red}{$\mathbf{0.48\pm0.53}$} & \textcolor{red}{$\mathbf{0.64\pm0.34}$} & \textcolor{red}{$\mathbf{78.20\pm1.35}$} & \textcolor{red}{$\mathbf{0.37\pm0.22}$} & \textcolor{red}{$\mathbf{0.41\pm0.39}$} \\
 \bottomrule
 \toprule
 Missing Rate & \multicolumn{3}{c}{\textbf{AMiner-S(0.1)}} & \multicolumn{3}{c}{\textbf{AMiner-S(0.2)}} & \multicolumn{3}{c}{\textbf{AMiner-S(0.3)}} \\
 Metrics & ACC(\%)$_{\uparrow}$ & $\Delta_\text{SP}$(\%)$_{\downarrow}$ & $\Delta_\text{EO}$(\%)$_{\downarrow}$ & ACC(\%)$_{\uparrow}$ & $\Delta_\text{SP}$(\%)$_{\downarrow}$ & $\Delta_\text{EO}$(\%)$_{\downarrow}$ & ACC(\%)$_{\uparrow}$ & $\Delta_\text{SP}$(\%)$_{\downarrow}$ & $\Delta_\text{EO}$(\%)$_{\downarrow}$\\
 \midrule
 GCN & $88.85\pm1.09$ & $5.19\pm2.63$ & $4.98\pm1.34$ & $83.72\pm1.96$ & $8.61\pm2.09$ & $8.58\pm2.12$ & $83.13\pm1.32$ & $7.12\pm1.16$ & $6.43\pm1.09$ \\
 SGFormer & \textcolor{red}{$\mathbf{89.54\pm0.07}$} & $4.63\pm0.69$ & $4.75\pm0.25$ & $88.30\pm0.08$ & $4.64\pm0.05$ & $5.18\pm1.61$ & \textcolor{blue}{\underline{$87.88\pm0.18$}} & $4.33\pm0.06$ & $3.31\pm0.53$ \\
 CoBFormer & $88.13\pm1.26$ & $4.65\pm0.94$ & $5.36\pm2.65$ & $87.86\pm0.28$ & $4.46\pm0.78$ & $4.99\pm3.60$ & $86.26\pm0.40$ & $5.19\pm1.09$ & \textcolor{blue}{\underline{$3.05\pm1.57$}} \\
 Polynormer & \textcolor{blue}{\underline{$89.42\pm0.19$}} & $5.16\pm0.91$ & $3.98\pm1.51$ & \textcolor{red}{$\mathbf{88.68\pm0.34}$} & $4.11\pm0.21$ & $5.86\pm2.70$ & \textcolor{red}{$\mathbf{87.99\pm1.34}$} & $4.31\pm0.79$ & $4.18\pm1.79$ \\
 FMP & $88.33\pm0.96$ & $7.10\pm2.02$ & $4.33\pm1.92$ & $86.46\pm1.52$ & $5.81\pm1.57$ & $6.03\pm1.63$ & $85.84\pm1.26$ & $4.44\pm1.65$ & $4.24\pm0.78$ \\
 FairSIN & $85.74\pm1.60$ & $4.85\pm3.78$ & $4.88\pm1.72$ & $84.25\pm0.01$ & $4.52\pm1.89$ & \textcolor{blue}{\underline{$4.68\pm1.24$}} & $84.04\pm0.16$ & $4.29\pm0.24$ & $3.28\pm0.22$ \\
 FairGT & $89.05\pm1.00$ & $4.51\pm0.60$ & $4.05\pm0.89$ & $87.25\pm1.11$ & \textcolor{blue}{\underline{$3.54\pm0.23$}} & $5.88\pm1.90$ & $86.59\pm0.42$ & $3.60\pm0.63$ & $3.57\pm1.95$ \\
 FairGNN & $86.00\pm1.05$ & $4.03\pm1.63$ & \textcolor{blue}{\underline{$3.75\pm2.86$}} & $87.59\pm1.81$ & $4.81\pm2.39$ & $5.44\pm2.37$ & $86.62\pm1.11$ & $4.48\pm0.65$ & $4.23\pm1.26$ \\
 FairAC & $78.94\pm2.04$ & \textcolor{blue}{\underline{$3.63\pm2.10$}} & $6.37\pm2.03$ & $79.61\pm0.02$ & $3.93\pm0.04$ & $6.43\pm1.31$ & $80.86\pm0.06$ & \textcolor{blue}{\underline{$3.47\pm1.13$}} & $5.88\pm1.51$ \\
 \textbf{FairGE} & $89.12\pm1.34$ & \textcolor{red}{$\mathbf{3.59\pm0.28}$} & \textcolor{red}{$\mathbf{2.97\pm0.37}$} & \textcolor{blue}{\underline{$88.33\pm2.03$}} & \textcolor{red}{$\mathbf{3.52\pm1.05}$} & \textcolor{red}{$\mathbf{3.52\pm1.06}$} & $87.01\pm1.01$ & \textcolor{red}{$\mathbf{2.94\pm0.49}$} & \textcolor{red}{$\mathbf{1.69\pm0.54}$} \\
 \midrule
 \midrule
 Missing Rate & \multicolumn{3}{c}{\textbf{AMiner-S(0.4)}} & \multicolumn{3}{c}{\textbf{AMiner-S(0.5)}} & \multicolumn{3}{c}{\textbf{AMiner-S(0.6)}} \\
 \midrule
 GCN & $83.08\pm1.87$ & $5.81\pm1.27$ & $4.96\pm1.77$ & $81.48\pm1.36$ & $4.49\pm0.90$ & $5.67\pm1.08$ & $81.89\pm1.71$ & $4.41\pm1.01$ & $6.13\pm1.46$ \\
 SGFormer & \textcolor{blue}{\underline{$85.54\pm1.35$}} & $4.87\pm0.86$ & $4.11\pm2.07$ & \textcolor{blue}{\underline{$84.57\pm0.51$}} & $4.24\pm0.37$ & $4.00\pm1.74$ & $83.47\pm1.00$ & $4.90\pm0.42$ & $4.06\pm0.84$ \\
 CoBFormer & $84.84\pm1.29$ & \textcolor{blue}{\underline{$2.54\pm0.33$}} & $4.40\pm0.64$ & $83.32\pm2.58$ & \textcolor{blue}{\underline{$2.19\pm0.19$}} & $5.46\pm2.29$ & \textcolor{blue}{\underline{$84.18\pm0.30$}} & $3.49\pm0.14$ & $4.07\pm3.97$ \\
 Polynormer & $85.04\pm2.48$ & $3.79\pm2.01$ & $3.40\pm1.49$ & \textcolor{red}{$\mathbf{85.14\pm1.94}$} & $3.57\pm1.69$ & $4.11\pm1.91$ & $83.17\pm0.09$ & $3.41\pm0.14$ & $4.87\pm0.75$ \\
 FMP & $85.42\pm1.25$ & $4.76\pm0.17$ & $3.90\pm0.17$ & $84.23\pm1.65$ & $4.47\pm1.16$ & $4.76\pm0.78$ & \textcolor{red}{$\mathbf{85.50\pm1.32}$} & $3.80\pm0.62$ & $5.68\pm0.69$ \\
 FairSIN & $82.15\pm0.07$ & $3.16\pm0.04$ & \textcolor{blue}{\underline{$3.15\pm0.04$}} & $80.06\pm0.29$ & $2.21\pm0.31$ & \textcolor{blue}{\underline{$2.07\pm0.05$}} & $79.91\pm0.80$ & $3.63\pm2.32$ & \textcolor{blue}{\underline{$3.69\pm3.46$}} \\
 FairGT & $84.26\pm1.88$ & $2.66\pm1.39$ & $4.65\pm2.04$ & $82.12\pm0.38$ & $2.60\pm0.61$ & $4.23\pm1.87$ & $80.24\pm0.00$ & $2.94\pm0.60$ & $3.92\pm0.42$ \\
 FairGNN & $83.58\pm0.35$ & $4.84\pm1.98$ & $4.63\pm2.49$ & $83.36\pm1.77$ & $4.75\pm1.22$ & $4.64\pm2.54$ & $82.58\pm1.93$ & \textcolor{blue}{\underline{$2.59\pm1.28$}} & $6.33\pm1.18$ \\
 FairAC & $81.44\pm0.25$ & $2.81\pm0.35$ & $5.92\pm1.40$ & $81.44\pm1.61$ & $4.18\pm1.91$ & $11.09\pm2.01$ & $81.93\pm0.74$ & $5.23\pm2.95$ & $10.14\pm2.74$ \\
 \textbf{FairGE} & \textcolor{red}{$\mathbf{85.56\pm0.53}$} & \textcolor{red}{$\mathbf{2.26\pm1.89}$} & \textcolor{red}{$\mathbf{2.42\pm0.24}$} & $83.77\pm0.69$ & \textcolor{red}{$\mathbf{2.10\pm0.19}$} & \textcolor{red}{$\mathbf{1.74\pm0.01}$} & $83.52\pm0.35$ & \textcolor{red}{$\mathbf{1.92\pm0.29}$} & \textcolor{red}{$\mathbf{2.21\pm0.43}$} \\
\bottomrule
\end{tabular}
\label{tab:result_bail and credit}
\end{table*}

\section{Experiments}
\subsection{Datasets}
\par Node classification is adopted as the downstream task, employing seven real-world  social network datasets: 
\textbf{Facebook}~ \cite{leskovec2012learning}, \textbf{Income}~\cite{asuncion2007uci}, \textbf{Bail}~\cite{jordan2015effect}, \textbf{Credit}~\cite{yeh2009the}, \textbf{Pokec-z}~\cite{takac2012data}, \textbf{Pokec-n}~\cite{takac2012data}, and \textbf{AMiner-S}~\cite{wan2019aminer}.
These datasets cover diverse social contexts, including online friendship networks, financial and judicial social interactions, and academic collaboration.
Notably, \textbf{AMiner-S} includes multi-class sensitive attributes, where sensitive attributes span more than two distinct categories (e.g., multiple regions instead of binary gender).

\par To support fairness analysis across different types of sensitive attributes, the \textbf{Pokec} datasets are further divdided into variants: \textbf{Pokec-z-R} and \textbf{Pokec-n-R} use living region as the sensitive attribute, while \textbf{Pokec-z-G} and \textbf{Pokec-n-G} use gender.
For datasets with more than two ground-truth label classes, label $0$ and $1$ are reained, and all class greater than $1$ are merged into class $1$.
The data is split into $25$\% for validation and $25$\% for testing. 
For training, $100$ nodes are sampled from the \textbf{Facebook} dataset, $6,000$ from \textbf{Credit}, and $1,000$ from each of \textbf{Income}, \textbf{Bail}, \textbf{Pokec-z}, \textbf{Pokec-n}, and \textbf{AMiner-S}.
This parameter setting follows prior work in fairness-aware graph learning~\cite{dai2023learning, luo2024fugnn}.

\subsection{Baselines}
\par To evaluate FairGE, the study includes a comprehensive set of baselines.
These comprise one vanilla GNN, GCN~\cite{kipf2017semi}, and three state-of-the-art GTs: SGFormer~\cite{wu2023sgformer}, CoBFormer~\cite{xing2024less}, and Polynormer~\cite{deng2024polynormer}. 
For fairness-aware models, the benchmark incorporates two fairness-aware GNNs, FMP~\cite{jiang2024chasing} and FairSIN~\cite{yang2024fairsin}, alongside one fairness-aware GT, FairGT~\cite{luo2024fairgt}. 
Additionally, two attribute-generation capable fairness-aware GNNs, FairGNN~\cite{dai2023learning} and Fair-AC~\cite{guo2023fair}, are included for comparison.
The experimental setting is shown in Appendix C.

\begin{table*}[t]
\centering
\footnotesize
\caption{Ablation study on different missing rates on Facebook.}
\tabcolsep=0.25cm
\renewcommand{\arraystretch}{0.88}
\begin{tabular}{l|ccc|ccc|ccc}
\toprule
Missing Rate & \multicolumn{3}{c}{\textbf{Facebook(0.1)}} & \multicolumn{3}{c}{\textbf{Facebook(0.2)}} & \multicolumn{3}{c}{\textbf{Facebook(0.3)}} \\
Metrics & ACC(\%)$_{\uparrow}$ & $\Delta_\text{SP}$(\%)$_{\downarrow}$ & $\Delta_\text{EO}$(\%)$_{\downarrow}$ & ACC(\%)$_{\uparrow}$ & $\Delta_\text{SP}$(\%)$_{\downarrow}$ & $\Delta_\text{EO}$(\%)$_{\downarrow}$ & ACC(\%)$_{\uparrow}$ & $\Delta_\text{SP}$(\%)$_{\downarrow}$ & $\Delta_\text{EO}$(\%)$_{\downarrow}$\\
\midrule
FairGE & $88.17\pm1.63$ & $0.89\pm0.04$ & $1.12\pm0.15$ & $86.64\pm1.69$ & $0.89\pm0.17$ & $0.17\pm0.08$ & $85.50\pm0.99$ & $1.35\pm0.18$ & $0.95\pm0.29$ \\
~FairGE (1) & $85.50\pm1.98$ & $1.66\pm0.81$ & $2.15\pm0.59$ & $85.11\pm2.03$ & $1.37\pm0.39$ & $1.90\pm0.84$ & $83.59\pm2.20$ & $1.39\pm0.27$ & $1.98\pm0.35$ \\
~FairGE (Mean) & $87.79\pm1.56$ & $1.02\pm0.12$ & $1.98\pm0.32$ & $85.02\pm1.65$ & $1.45\pm0.34$ & $1.81\pm0.86$ & $81.68\pm1.43$ & $2.11\pm0.38$ & $2.16\pm0.54$ \\
~FairGE w/o ST & $82.06\pm2.11$ & $2.23\pm0.84$ & $2.08\pm0.23$ & $81.68\pm1.09$ & $1.94\pm0.16$ & $1.46\pm0.32$ & $79.39\pm2.21$ & $3.51\pm1.54$ & $2.88\pm1.07$ \\
\midrule
\midrule
Missing Rate & \multicolumn{3}{c}{\textbf{Facebook(0.4)}} & \multicolumn{3}{c}{\textbf{Facebook(0.5)}} & \multicolumn{3}{c}{\textbf{Facebook(0.6)}} \\
\midrule
FairGE & $84.73\pm1.97$ & $0.75\pm0.19$ & $1.38\pm0.89$ & $83.59\pm1.73$ & $0.73\pm0.49$ & $0.95\pm0.38$ & $83.59\pm1.83$ & $1.27\pm0.68$ & $0.17\pm0.04$ \\
~FairGE (1) & $80.15\pm0.89$ & $1.03\pm0.12$ & $0.21\pm1.45$ & $82.44\pm0.32$ & $1.63\pm0.67$ & $1.64\pm0.45$ & $76.34\pm0.23$ & $1.45\pm0.76$ & $2.15\pm0.21$ \\
~FairGE (Mean) & $80.15\pm1.56$ & $0.91\pm0.08$ & $2.41\pm1.23$ & $77.86\pm1.56$ & $1.78\pm0.45$ & $1.90\pm0.67$ & $75.57\pm2.34$ & $1.90\pm0.87$ & $2.34\pm0.45$ \\
~FairGE w/o ST & $77.86\pm1.23$ & $3.99\pm0.34$ & $1.91\pm0.98$ & $76.33\pm1.67$ & $2.48\pm0.56$ & $2.68\pm0.23$ & $76.72\pm2.12$ & $3.50\pm0.45$ & $2.42\pm0.67$ \\
\bottomrule
\end{tabular}
\label{tab:ablation}
\end{table*}

\begin{table}[t]
    \centering
    \small
    \caption{Runtime (s) of GT-based baseline models across seven datasets. OOM indicates out-of-memory errors.}
    \tabcolsep=0.08cm
    \renewcommand{\arraystretch}{0.85}
    \begin{tabular}{lccccc}
        \toprule
        \textbf{Dataset} & \textbf{FairGE} & \textbf{SGFormer} & \textbf{CoBFormer} & \textbf{Polynormer} & \textbf{FairGT} \\
        \midrule
        \textbf{Facebook} & \textcolor{red}{$12.96$} & $12.99$ & $17.19$ & $14.74$ & $13.53$  \\
        \textbf{Income} & \textcolor{red}{$\mathbf{19.77}$} & $24.97$ & $25.58$ & $22.65$ & $22.83$ \\
        \textbf{Bail} & \textcolor{red}{$\mathbf{18.89}$} & $33.09$ & $28.06$ & $26.41$ & $24.56$ \\
        \textbf{Credit} & \textcolor{red}{$\mathbf{23.15}$} & $45.70$ & $31.92$ & $29.89$ & $31.34$ \\
        \textbf{Pokec-z} & \textcolor{red}{$\mathbf{18.94}$} & OOM & $39.01$ & $34.13$ & $39.16$ \\
        \textbf{Pokec-n} & \textcolor{red}{$\mathbf{18.13}$} & OOM & $37.21$ & $31.94$ & $37.81$ \\
        \textbf{AMiner-S} & \textcolor{red}{$\mathbf{15.69}$} & OOM & $22.42$ & $18.91$ & $29.98$ \\
        \bottomrule
    \end{tabular}
    \label{tab:cost time}
\end{table}

\subsection{Comparison Results}
\par To simulate incomplete sensitive features, this study uniformly masks values in the complete sensitive attribute column. 
This methodology is consistently applied across all models to ensure fair comparison, with both original and incomplete-data distributions explicitly defined in the experiments.
Figure~\ref{fig:results_sp} and Figure~\ref{fig:results_eo} provides a detailed comparison of the fairness evaluation metrics between the proposed FairGE and several baseline models across seven social network datasets with varying missing rates.
Specifically, comparison experiments are conducted with missing rates ranging from $10\%$ to $60\%$, covering both sparse and dense incompleteness scenarios.
The overall accuracy, along with two key fairness metrics, $\Delta_{\text{SP}}$ and $\Delta_{\text{EO}}$, part of them are reported in Table~\ref{tab:result_bail and credit}, and others are shown in \url{https://github.com/LuoRenqiang/FairGE}.

\par The \textbf{AMiner-S} dataset contains multi-class sensitive attributes.
While previous fairness evaluation metrics are formulated for binary scenarios, they can be readily extended to multi-class sensitive attribute setting, as demonstrated by previous work~\cite{luo2025fairgp}. 
To ensure fairness across all sensitive subgroups, we quantify the core principle using the variance between them.
Specifically, the fairness evaluation metrics for multi-class sensitive attributes are defined as follows, where class number denotes $m$:
\begin{equation}
    \begin{aligned}        
    &\Delta_\text{SP}=\text{Var}_{i=1}^m\big(|\mathbb{P}(\hat{y}=1|s=i)\big), \\        &\Delta_\text{EO}=\text{Var}_{i=1}^m\big(|\mathbb{P}(\hat{y}=1|y=1,s=i)\big),
    \end{aligned}
\end{equation}

\par FairGE consistently achieves superior fairness in node classification, outperforming other GTs and fairness-aware GNNs, even those generating data, especially under high incomplete data rates, a challenge where other methods like FairGT struggle. 
Beyond its robust fairness, FairGE also maintains high accuracy across all seven social network datasets. 
This exceptional performance extends to handling multiple sensitive attributes simultaneously, as evidenced by experiments on multi-sensitive datasets (Pokec-z and Pokec-n), where FairGE not only enhances fairness for individual sensitive attributes but also maintains it consistently across multiple ones, solidifying its role as a comprehensive fairness-aware solution for GTs with incomplete sensitive attributes.

\subsection{Ablation Study}
\par An ablation study reveals the individual impact of these techniques on prediction fairness and accuracy. 
The study systematically removed each component; for instance, the version without spectral truncation $\textbf{w/o ST}$ was compared against the full FairGE model. 
Results in Table~\ref{tab:ablation} consistently show FairGE outperforming $\textbf{w/o ST}$ across accuracy, statistical parity ($\Delta_{\text{SP}}$), and equal opportunity ($\Delta_{\text{EO}}$ ), emphasizing spectral truncation's crucial role in preserving key data characteristics and ensuring fair outcomes.

\par Furthermore, the study analyzed the impact of zero padding by comparing FairGE with variants using different imputation strategies: FairGE (1), which pads with '1', and FairGE (Mean), which fills with attribute means. 
These comparisons aimed to understand how different incomplete attribute handling strategies affect model performance and fairness. 
The results confirmed that zero padding leads to significant improvements, as FairGE consistently achieved higher accuracy and lower fairness disparity ($\Delta_{\text{SP}}$ and $\Delta_{\text{EO}}$ ) compared to these ablated versions. 
This reinforces the importance of zero padding in enhancing both fairness and accuracy. 
In summary, the ablation study underscores the essential roles of both spectral truncation and zero padding in improving the fairness and performance of GTs within FairGE.
 
\subsection{Training Cost Comparison}
\par FairGE employs spectral truncation for encoding, which offers efficiency advantages over standard GTs. 
To validate this, we compare its training time with various GT baselines. 
For fairness, all methods are standardised with $128$ hidden dimensions, $1$ layer, $1$ attention head, and $500$ training epochs. 
As shown in Table~\ref{tab:cost time}, FairGE achieves superior accuracy and fairness in node classification without additional computational cost. 
This demonstrates its scalability, making FairGE a robust and efficient framework for fairness-aware GTs.

\section{Conclusion}
\par This paper addresses fairness issues in incomplete social networks. 
The proposed framework, FairGE, introduces two key components: zero-padding to handle incomplete sensitive attributes and spectral truncation to preserve multi-hop neighbor information. 
These components operate together to mitigate bias and maintain the independence of sensitive attributes in incomplete social networks. 
The design of FairGE enables effective encoding of both node and structural information, even in the absence of sensitive data, while avoiding synthetic attribute generation and thereby alleviating privacy concerns. 
Theoretical analysis demonstrates that the combination of spectral truncation and zero-padding maximises fairness while preserving model utility. 
Extensive experiments on seven social network datasets demonstrate that FairGE significantly improves fairness compared with state-of-the-art models, achieving substantial gains in statistical parity and equality of opportunity. 
Despite these strong results, FairGE remains sensitive to parameter selection (e.g., the number of principal eigenvectors and the padding strategy), highlighting the need for further optimisation. 
Future work will aim to enhance efficiency, robustness, and applicability to a broader range of sensitive attribute settings.

\bibliographystyle{ACM-Reference-Format}
\bibliography{fairge}

\appendix
\section{Ethical Use of Data and Informed Consent}
All datasets used in this study (e.g., Credit, AMiner-S, and other benchmark social network datasets) are publicly available under their respective research licenses. 
These datasets were released for research purposes and have been widely adopted in prior work on graph learning and fairness-aware machine learning. 
They do not contain personally identifiable information beyond what has been made publicly accessible, and no additional data collection or annotation was conducted by the authors. 
As this work relies solely on secondary analysis of existing open datasets, no direct interaction with human participants occurred, and informed consent was not required. 
The study complies with the ethical use of data guidelines established by the research community.

\section{Theoretical Analysis and Proofs}
\label{Proofs}
\textbf{Theorem 1}: \textit{The effect of utilising the principal eigenvector as a structural encoding, in comparison to using multiple adjacency matrices as structural topology encoding for multi-hop sensitive attributes under incomplete sensitive attributes, can be approximated.
The following equation holds:}
    \begin{equation} 
        \lim_{k \to \infty} \cos(\langle \mathbf{A}^k \mathbf{H}'[:,s], \mathbf{H}'[:,s] \rangle) = \cos(\langle \mathbf{p}_1, \mathbf{H}'[:,s] \rangle). 
    \end{equation}

\textit{Proof.}
We assume $\gamma_i = \textbf{H}'[:,s]^\top \textbf{p}_i$.
Since $\mathbf{A} \in \mathbb{R}^{n \times n}$ is a symmetric matrix, the eigendecomposition of $\mathbf{A}$ can be written as $\mathbf{P} = (\mathbf{p}_1, \mathbf{p}_2, \dots, \mathbf{p}_n)$, and $\Vert \mathbf{p}_i \Vert = 1$ ($i \in \{1, 2, \dots, n\}$) and $\Lambda = diag(\lambda_1, \lambda_2, \dots, \lambda_n)$. 
Thus,
\begin{equation*}
    \begin{aligned}
        &\lim_{k \to \infty} cos(\langle\mathbf{A}^k \textbf{H}'[:,s], \textbf{H}'[:,s]\rangle) \\
        =&\lim_{l\rightarrow\infty}\frac{(\mathbf{A}^k \textbf{H}'[:,s])^\top \textbf{H}'[:,s]}{\Vert \mathbf{A}^k \textbf{H}'[:,s] \Vert \Vert\textbf{H}'[:,s]  \Vert}\\
        =&\lim_{l\rightarrow\infty}\frac{(\mathbf{A}^k \textbf{H}'[:,s])^\top \textbf{H}'[:,s]}{\sqrt{(\mathbf{A}^k \textbf{H}'[:,s])^\top \mathbf{A}^k \textbf{H}'[:,s]} \sqrt{\textbf{H}'[:,s]^\top \textbf{H}'[:,s]}}.
    \end{aligned}    
\end{equation*}

Because,
\begin{equation*}
    \begin{aligned}
    &\mathbf{A}^k \textbf{H}'[:,s] = \mathbf{P} \Lambda^k \mathbf{P}^\top \mathbf{H}'[:,s], \\
    &\gamma_i = \textbf{H}'[:,s]^\top \textbf{p}_i.
    \end{aligned}    
\end{equation*}

Thus,
\begin{equation*}
    \begin{aligned}
        &\lim_{k \to \infty} cos(\langle\mathbf{A}^k \textbf{H}'[:,s], \textbf{H}'[:,s]\rangle) \\
        =&\lim_{k \to \infty} \frac{\sum^n_{i=1} \gamma_i^2 \lambda_i^k}{\sqrt{\sum^n_{i=1}\gamma_i^2 \lambda_i^{2k}}\sqrt{\sum^n_{i=1}\gamma_i^2}}\\
        =&\lim_{k \to \infty} \frac{\gamma_1^2 + \sum^n_{i=2}\gamma_i^2 (\frac{\lambda_i}{\lambda_1})^k}{\sqrt{\gamma_1^2 + \sum^n_{i=2}\gamma_i^2 (\frac{\lambda_i}{\lambda_1})^{2k}}\sqrt{\sum^n_{i=1}\gamma_i^2}}\\
        =&\frac{\gamma_1}{\sqrt{\sum^n_{i=1}\gamma_i^2}}.
    \end{aligned}    
\end{equation*}    

Because,
\begin{equation*}
    \begin{aligned}
        cos(\langle \mathbf{p}_i, \mathbf{H}'[:, s] \rangle) =& \frac{\mathbf{H}'[:,s]^\top \mathbf{p}_i}{\Vert \mathbf{H}'[:,s] \Vert \Vert \mathbf{p}_i \Vert} = \frac{\mathbf{H}'[:.s]^\top \mathbf{p}_i}{\Vert \mathbf{H}'[:.s] \Vert} \\ 
        =& \frac{\mathbf{H}'[:.s]^\top \mathbf{p}_i}{\sqrt{\mathbf{H}'[:.s]^\top \mathbf{H}'[:.s]}} \\
        =& \frac{\mathbf{H}'[:.s]^\top \mathbf{p}_i}{\sqrt{(\mathbf{P}^\top \mathbf{H}'[:.s])^\top \mathbf{P}^\top \mathbf{H}'[:.s]}} \\
        =& \frac{\mathbf{H}'[:.s]^\top \mathbf{p}_i}{\sqrt{\sum_{j=1}^n(\mathbf{H}'[:.s]^\top \mathbf{p}_j)^2}}\\
        =& \frac{\gamma_i}{\sqrt{\sum^n_{j=1}\gamma_j^2}}.\\
    \end{aligned}    
\end{equation*}

Then,
\begin{equation*}
    \lim_{k \to \infty} cos(\langle\mathbf{A}^k \textbf{H}'[:,s], \textbf{H}'[:,s]\rangle) = cos(\langle \mathbf{p}_1, \textbf{H}'[:,s] \rangle).
\end{equation*}

\rightline{$\square$}

\par \textbf{Theorem 1.1}: \textit{In cases where multiple maximal eigenvalues exist, the conclusions remain analogous, with further elaboration as following}:
\begin{equation}
    \begin{aligned}
        & \lim_{l \to \infty} cos(\langle\mathbf{A}^k \mathbf{H}'[:,s], \mathbf{H}'[:,s]\rangle) \\ 
        \geq &\frac{1}{\sqrt{j}}\sum \nolimits_{i=1}^j{cos(\langle \mathbf{H}'[:,s], \mathbf{p}_i\rangle)}.
    \end{aligned}
\end{equation}
\textit{Proof}. 
\begin{equation}
    \nonumber
    \begin{aligned}
	&{\lim_{k \rightarrow \infty}}cos(\langle\mathbf{A}^k \mathbf{H}'[:,s], \mathbf{H}'[:,s]\rangle) \\ =&\lim_{k\rightarrow\infty}\frac{\sum^n_{i=1}\gamma_i^2 \lambda_i^k}{\sqrt{\sum^n_{i=1}\gamma_i^2 \lambda_i^{2k}}\sqrt{\sum^n_{i=1}\gamma_i^2}} \\ 
    =&\lim_{k\rightarrow\infty}\frac{\sum^j_{i=1}\gamma_i^2 + \sum^n_{i=j+1}\gamma_i^2 (\frac{\lambda_i}{\lambda_1})^k}{\sqrt{\sum^j_{i=1}\gamma_i^2 + \sum^n_{i=j+1}\gamma_i^2 (\frac{\lambda_i}{\lambda_1})^{2k}}\sqrt{\sum^n_{i=1}\gamma_i^2}}\\
    =&\frac{\sqrt{\sum_{i=1}^j\gamma_i^2}}{\sqrt{\sum^n_{i=1}\gamma_i^2}}.
    \end{aligned}
\end{equation}

Then, considering $cos(\langle \mathbf{H}'[:,s], \mathbf{p}_i\rangle)=\frac{\gamma_i}{\sqrt{\sum^n_{i=1}\gamma_i^2}}$ and the Cauchy-Schwarz Inequality, we have:
\begin{equation}
    \nonumber
    \begin{aligned}
	&\lim_{l \to \infty} cos(\langle\mathbf{A}^k \mathbf{H}'[:,s], \mathbf{H}'[:,s]\rangle) = \frac{1}{\sqrt{j}}\frac{\sqrt{j*1^2\sum_{i=1}^j\gamma_i^2}}{\sqrt{\sum^n_{i=1}\gamma_i^2}} \\
    \geq& \frac{1}{\sqrt{j}}\frac{\sum_{i=1}^j\gamma_i}{{\sqrt{\sum_{i=1}^n\gamma_i^2}}} = \frac{1}{\sqrt{j}}\sum \nolimits_{i=1}^j{cos(\langle \mathbf{H}'[:,s], \mathbf{p}_i\rangle)}.
    \end{aligned}
\end{equation}

Especially, only if right-side terms $cos(\langle \mathbf{H}'[:,s], \mathbf{p}_i\rangle)$ are equal, the equation holds.

\rightline{$\square$}

\par \textbf{Theorem 1.2} \textit{The effect of non-principal eigenvalues diminishes at an exponential rate.}

\textit{Proof}.
Let $\mathbf{T}^k = cos(<\mathbf{A}^k \mathbf{H}'[:,s], \mathbf{H}'[:,s])$, thus:
\begin{equation*}
  \mathbf{T}^k = \frac{\gamma_1^2+\sum_{i=2}^n \gamma_i^2 (\frac{\lambda_i}{\lambda_1})^k}{\sqrt{\gamma_1^2+\sum_{i=2}^n \gamma_i^2 (\frac{\lambda_i}{\lambda_1})^{21}} \cdot \sqrt{\sum_{i=1}^n \gamma_i^2}}.
\end{equation*}

Let $\frac{\gamma_i}{\gamma_1}=u_i$, $\frac{\lambda_i}{\lambda_1}=v_i<1$, thus:
\begin{equation*}
\mathbf{T}^k=\frac{1+\sum_{i=2}^n u_i^2 v_i^k}{\sqrt{1+\sum_{i=2}^n u_i^2 v_i^{2 l}} \sqrt{1+\sum_{i=2}^n u_i^2}}.
\end{equation*}

Let $ \mathbf{T}'=\lim _{l \rightarrow \infty} \mathbf{T}^k$, thus:
\begin{equation*}
  \mathbf{T}' =\frac{\gamma_1}{\sqrt{\sum_{i=1}^n \gamma_i^2}}=\frac{1}{\sqrt{1+\sum_{i=2}^n u_i^2}}.
\end{equation*} 

Using the Taylor expansion of the multivariate function $f(\mathbf{x})$ (second order):
\begin{equation*}
  \begin{aligned}
f(\mathbf{x})= & f(\mathbf{x}_0)+\nabla f(\mathbf{x}_0)^{\top}(\mathbf{x}-\mathbf{x}_0) \\ +&\frac{1}{2}(\mathbf{x}-\mathbf{x}_0)^{\top} \nabla^2 f(\mathbf{x}_0)(\mathbf{x}-\mathbf{x}_0) +o(||\mathbf{x}-\mathbf{x}_0||)^2
  \end{aligned}
\end{equation*}

Let $x_i=u_i v_i^k$, $ f(x_2, \ldots, x_n)=\sqrt{1+\sum_{i=2}^n x_i^2}$.
Thus,
\begin{equation*}
  \frac{\partial f}{\partial x_i}=\frac{1}{2} \cdot \frac{1 \cdot 2 x_i}{\sqrt{1+\sum_{i=2}^n x_i^2}}=\frac{x_i}{\sqrt{1+\sum_{i=2}^n x_i^2}}.
\end{equation*}

When $x_2 = \ldots = x_n = 0$, 
\begin{equation*}
    \frac{\partial f}{\partial x_i} = 0
\end{equation*}
\begin{equation*}
    \begin{aligned}
        &\frac{\partial f}{\left(\partial x_i\right)^2} \\
        =&\frac{\sqrt{1+\sum_{i=2}^n x_i^2} \cdot 1-x_i \cdot\left(-\frac{1}{2}\right) \cdot \frac{2 x_i}{\sqrt{\left(1+\sum_{i=2}^n x_i^2\right)^3}}}{1+\sum_{i=2}^n x_i^2} \\
        =&\frac{1}{\sqrt{1+\sum_{i=2}^n x_i^2}}+\frac{x_i^2}{\sqrt{(1+\sum_{i=2}^n x_i^2)^5}}.
    \end{aligned}
\end{equation*}

When $x_2 = \ldots = x_n = 0$, 
\begin{equation*}
    \frac{\partial f}{\left(\partial x_i\right)^2} = 1.    
\end{equation*}
\begin{equation*}
 \frac{\partial f}{\partial x_i \partial x_j}=\frac{x_i x_j}{\sqrt{\left(1+\sum_{i=2}^n x_i{ }^2\right)^5}}. 
\end{equation*}

When $x_2 = \ldots = x_n = 0$, 
\begin{equation*}
    \frac{\partial f}{\partial x_i \partial x_j} = 0.    
\end{equation*}

Thus, the Taylor expansion of $f(x_2, \ldots, x_n)$ at $(0, \ldots, 0)$:
\begin{equation*}
  \begin{aligned}
  f\left(x_2, \ldots, x_n\right) & =f(0, \ldots, 0)+\sum_{i=2}^n\left(x_i-0\right) f^{\prime} x_i(0, \ldots, 0)  \\
  &+\frac{1}{2} \sum_{i, j=2}^n\left(x_i-0\right)\left(x_j-0\right) f^{\prime \prime} x_i x_j(0, \ldots, 0)  \\
  &+o^2 \\ 
  & =1+0+\frac{1}{2} \sum_{i=2}^n x_i^2+o^2 \\
  & =1+\frac{1}{2} \sum_{i=2}^n x_i^2+o^2.
  \end{aligned}
\end{equation*}

Thus:
\begin{equation*}
  \sqrt{1+\sum_{i=2}^n\left(u_i v_i^k\right)^2}=1+\frac{1}{2} \sum_{i=2}^n\left(u_i v_i^k\right)^2+o^2, 
\end{equation*}
\begin{equation*}
  \sqrt{1+\sum_{i=2}^n\left(u_i v_i^{l+1}\right)^2}=1+\frac{1}{2} \sum_{i=2}^n\left(u_i v_i^{l+1}\right)^2+o^2.
\end{equation*}

Because $\lim _{l \rightarrow \infty}\left(1+\frac{1}{2} \sum_{i=2}^n\left(u_i v_i^k\right)^2\right)=1$, thus
\begin{equation*}
  \begin{aligned}
  \lim _{l \rightarrow \infty} \frac{\mathbf{T}^{l+1}-\mathbf{T}'}{\mathbf{T}^k-\mathbf{T}'}&=\lim _{l \rightarrow \infty} \frac{\sum_{i=2} u_i^2 v_i^{l+1}-\sum_{i=2} \frac{1}{2} u_i^2 v_i^{2l+2}}{\sum_{i=2}^n u_i^2 v_i^k-\sum_{i=2}^n \frac{1}{2} u_i^2 v_i^{2 l}} \\
  & =\lim _{l \rightarrow \infty} \frac{\sum_{i=2}^n u_i^2\left(v_i^{l+1}-\frac{1}{2} v_i^{2 l+2}\right)}{\sum_{i=2}^n u_i^2\left(v_i^k-\frac{1}{2} v_i^{2 l}\right)} \\
  & =\lim _{l \rightarrow \infty} \frac{\sum_{i=2}^n u_i^2 v_i^{l+1}}{\sum_{i=2}^n u_i^2 v_i^k} \\
  & =\lim _{l \rightarrow \infty} \frac{u_2^2 v_2^{l+1}+\sum_{i=3}^n u_i v_i^{l+1}}{u_2^2 v_2^k+\sum_{i=3}^n u_i v_i^k}. \\
  &
  \end{aligned}
\end{equation*}

Let $\frac{v_i}{v_2}=\omega_i < 1$, thus:
\begin{equation*}
  \begin{aligned}
  \lim _{l \rightarrow \infty} \frac{\mathbf{T}^{l+1}-\mathbf{T}'}{\mathbf{T}^k-\mathbf{T}'} & =\lim _{l \rightarrow \infty} \frac{u_2^2 v_2 \cdot v_2^k+\sum_{i=3}^n u_i \omega_i^k v_i \cdot v_2^k}{u_2^2 \cdot v_2^k+\sum_{i=3}^n u_i w_i^k \cdot v_2^k} \\
  & =\lim _{l \rightarrow \infty} \frac{u_2^2 v_2+\sum_{i=3}^n u_i w_i^k v_i}{u_2^2+\sum_{i=3}^n u_i \omega_i^k} \\
  & =v_2 \\
  & =\frac{\lambda_2}{\lambda_1}.
  \end{aligned}
\end{equation*}

\par Therefore, $cos(<\mathbf{A}^k \mathbf{H}'[:,s], \mathbf{H}'[:,s]>)$ decays exponentially with the rate determined by $\frac{\lambda_2}{\lambda_1}$.

\rightline{$\square$}

\textbf{Theorem 2}: \textit{When sensitive attributes are incomplete, utilizing principal eigenvectors as structural topology encoding can effectively replicate the results achieved by using the adjacency matrix when the sensitive attributes are complete. 
The following equation holds:}
\begin{equation} 
    \lim_{k \to \infty} \cos(\langle \mathbf{A}^k \mathbf{H}[:,s], \mathbf{H}'[:,s] \rangle) = \cos(\langle \mathbf{p}_1, \mathbf{H}'[:,s] \rangle). 
\end{equation}

\textit{Proof.}
We also assume $\alpha_i = \textbf{H}[:.s]^\top \textbf{p}_i$ and $\gamma_i = \textbf{H}'[:.s]^\top \textbf{p}_i$.
Thus,
\begin{equation*}
    \begin{aligned}
        &\lim_{k \to \infty} cos(\langle\mathbf{A}^k \textbf{H}[:.s], \textbf{H}'[:.s]\rangle) \\
        =&\lim_{k \to \infty} \frac{\sum^n_{i=1}\alpha_i \gamma_i \lambda_i^k}{\sqrt{\sum^n_{i=1}\alpha_i^2 \lambda_i^{2k}}\sqrt{\sum^n_{i=1}\gamma_i^2}}\\
        =&\lim_{k \to \infty} \frac{\sum^n_{i=2}\alpha_i \gamma_i (\frac{\lambda_i}{\lambda_1})^k + \alpha_1 \gamma_1}{\sqrt{\alpha_1^2 + \sum^n_{i=2}\alpha_i^2 (\frac{\lambda_i}{\lambda_1})^{2k}}\sqrt{\sum^n_{i=1}\gamma_i^2}}\\
        =&\frac{\gamma_1}{\sqrt{\sum^n_{i=1}\gamma_i^2}}.
    \end{aligned}    
\end{equation*}

Because,
\begin{equation*}
    cos(\langle \mathbf{p}_1, \mathbf{H}'[:, S] \rangle) = \frac{\gamma_1}{\sqrt{\sum^n_{i=1}\gamma_i^2}}.
\end{equation*}

Then,
\begin{equation*}
    \lim_{k \to \infty} cos(\langle\mathbf{A}^k \textbf{H}[:.s], \textbf{H}'[:.s]\rangle) = cos(\langle \mathbf{p}_1, \textbf{H}'[:.s] \rangle).
\end{equation*}

\rightline{$\square$}

\textbf{Theorem 3}: \textit{Padding the incomplete value with zeros can effectively recover the neighbouring sensitive attribute information in the scenario where sensitive attributes are incomplete. 
The following equation holds:}
\begin{equation}
    \begin{aligned}
        &\lim_{k \to \infty} cos(\langle\mathbf{A}^k \mathbf{H}'[:,s](0), \mathbf{H}[:,s]\rangle) \\
        =&\lim_{k \to \infty} cos(\langle\mathbf{A}^k \mathbf{H}[:,s], \mathbf{H}[:,s]\rangle).
    \end{aligned}
\end{equation}

\textit{Proof.}
We denote $\mathbf{H}'[:,s](0)$ as $\mathbf{H}'[:,s](0) = \mathbf{I}_{n,t} \mathbf{H}[:,s]$,
where 
\begin{equation*}
    \mathbf{I}_{n,t} = \left\{
        \begin{array}{ll}
        1, & i=j \quad \text{and} \quad v_i[s] \; \text{is complete},\\
        0, & \text{else}.
        \end{array}
        \right.
\end{equation*}

We assume $\alpha_i = \textbf{H}[:,s]^\top \textbf{p}_i$ and $\beta_i = \textbf{H}[:,s]^\top \textbf{I}_{n,t} \textbf{p}_i$.
Thus,
\begin{equation*}
    \begin{aligned}
        &||\textbf{A}^k\textbf{H}'[:,s](0)||^\top \\
        =&\sqrt{(\textbf{A}^k\textbf{H}'[:,s](0))^\top\textbf{A}^k\textbf{H}'[:,s](0)} \\
        =&\sqrt{\textbf{H}'[:,s](0)^\top \textbf{A}^{2k} \textbf{H}'[:,s](0)} \\
        =&\sqrt{\textbf{H}'[:,s](0)^\top \textbf{P} \Lambda^{2k} \textbf{P} \textbf{H}'[:,s](0)} \\
        =&\sqrt{\sum^n_{i=1}\beta_i^2 \lambda_i^{2k}}.
    \end{aligned}    
\end{equation*}
\begin{equation*}
        ||\textbf{H}[:,s]|| = \sqrt{(\textbf{H}[:,s])^\top\textbf{H}[:,s]} = \sqrt{\sum^n_{i=1}\alpha_i^2}.
\end{equation*}

Therefore,
\begin{equation*}
    \begin{aligned}
        &\lim_{k \to \infty} cos(\langle\mathbf{A}^k \textbf{H}'[:,s](0), \textbf{H}[:,s]\rangle) \\
        =&\lim_{k \to \infty} \frac{(\mathbf{A}^k \textbf{H}'[:,s](0))^\top \textbf{H}[:,s]}{||\mathbf{A}^k \textbf{H}'[:,s](0) ||^\top || \textbf{H}[:,s]||} \\
        =&\lim_{k \to \infty} \frac{\textbf{H}'[:,s]^\top (\textbf{I}_{n,t} \textbf{P}\Lambda^k\textbf{P}^\top)\textbf{H}[:,s]}{||\mathbf{A}^k \textbf{H}'[:,s](0) ||^\top || \textbf{H}[:,s]||} \\
        =&\lim_{k \to \infty} \frac{\sum^n_{i=1}\alpha_i \beta_i \lambda_i^k}{\sqrt{\sum^n_{i=1}\beta_i^2 \lambda_i^{2k}}\sqrt{\sum^n_{i=1}\alpha_i^2}}.\\
    \end{aligned}    
\end{equation*}

Because $\lambda_1 >> \lambda_i$:
\begin{equation*}
    \begin{aligned}
        &\lim_{k \to \infty} cos(\langle\mathbf{A}^k \textbf{H}'[:,s](0), \textbf{H}[:,s]\rangle) \\
        =&\lim_{k \to \infty} \frac{\alpha_1 \beta_1 + \sum^n_{i=2}\alpha_i \beta_i (\frac{\lambda_i}{\lambda_1})^k}{\sqrt{\beta_1^2 + \sum^n_{i=2}\beta_i^2 (\frac{\lambda_i}{\lambda_1})^{2k}}\sqrt{\sum^n_{i=1}\alpha_i^2}} \\
        =&\lim_{k \to \infty} \frac{\alpha_1 \beta_1}{\sqrt{\beta_1^2}\sqrt{\sum^n_{i=1}\alpha_i^2}}\\
        =&\frac{\alpha_1}{\sqrt{\sum^n_{i=1}\alpha_i^2}} \\
        =&\lim_{k \to \infty} cos(\langle\mathbf{A}^k \textbf{H}[:,s], \textbf{H}[:,s]\rangle).
    \end{aligned}    
\end{equation*}

\rightline{$\square$}

\section{Experimental Setting}
\label{Results}
\par All experiments are conducted on Ubuntu with $128$GB RAM and dual NVIDIA V$100$ GPUs ($16$GB each). 
The implementation is based on PyTorch $1$.$13$.$1$ and PyTorch-Geometric $2$.$5$.$3$. 
Model training uses the Adam optimizer~\cite{kingma2015adam} with $300$ epochs, a learning rate of $0.01$, and weight decay of $5e-4$. 
Hidden dimensions are varied across \{$16$, $32$, $64$, $128$\} for comprehensive evaluation. 
To test robustness under incomplete data, attribute missing rates are set to six levels: 
\{$0.1$, $0.2$, $0.3$, $0.4$, $0.5$, $0.6$\}. 
Model performance is assessed using accuracy for predictive capability, along with statistical parity and equality opportunity for fairness evaluation.

\end{document}